\documentstyle[pra,aps,multicol,psfig]{revtex}
\def\Figures{y}
\def\EprintServer{y}

\def\NarrowCaptions{y}

\def\TwoColumns{y}

\def\YES{y} % definition for TRUE

\if\TwoColumns\YES
  \def\narrowtextAAA{\begin{multicols}{2}}
  \def\widetextAAA{\end{multicols}  }
  \def\widetextEQ{\end{multicols}

     \noindent\vrule width 0.49\textwidth height 0.5pt\vrule width 0.5pt height 5pt\hfill

  }
  \def\narrowtextEQ{

    \noindent\hfill\vrule width 0.5pt height 0.5pt depth 4pt\vrule width 0.49\textwidth height 0.5pt depth 0pt

    \begin{multicols}{2}
  }
  \def\AppendixBEG{
     \begin{appendix}
     \begin{center} \rule{0.9\textwidth}{1pt} \end{center}
  }
  \def\AppendixEND{\end{appendix}}
\else
  \def\narrowtextAAA{\narrowtext}
  \def\widetextAAA{\widetext}
  \def\narrowtextEQ{\narrowtext}
  \def\widetextEQ{\widetext}
  \def\AppendixBEG{\begin{appendix}}
  \def\AppendixEND{\end{appendix}}
\fi

\def\gtsim{\gtrsim}
\def\ltsim{\lesssim}

\def\func#1{\mbox{#1}}

\def\vphantomA{\vphantom{ a^{(2)} }}

\def\vphantomAAA{\vphantom{\left[ \frac{\partial }{\partial a_{j}^{*}}\right] }}

  \if\NarrowCaptions\YES
    \def\widthAAA{3.1in} %captions
    \def\widthFIG{3.37in}
    \def\rwidthFIG{3.4in}
    \def\heightFIG{5.7in}
    \def\rheightFIG{6in}
    \def\widthFIGa{3.1in}
    \def\rwidthFIGa{3.2in}
  \else
    \widetextAAA
    \def\widthAAA{5.5in} %captions
    \def\widthFIG{5.5in} %figs
    \def\rwidthFIG{5.45in} %reserve so much
    \def\heightFIG{7in}
    \def\rheightFIG{6.5in}
    \def\widthFIGa{5.5in} %figs
    \def\rwidthFIGa{5.45in} %reserve so much
  \fi

\begin{document}
%\draft
\title{Cumulant expansion for studying damped quantum solitons}
\author{Eduard~Schmidt, Ludwig~Kn\"{o}ll, and Dirk--Gunnar~Welsch}
\address{Friedrich-Schiller-Universit\"{a}t Jena,
Theoretisch-Physikalisches Institut\\
Max-Wien-Platz 1, D-07743 Jena, Germany}
\maketitle

\begin{abstract}
The quantum statistics of damped optical solitons is studied using
cumulant-expansion techniques.
The effect of absorption is
described in terms of ordinary Markovian relaxation theory, by coupling the
optical field to a continuum of reservoir modes.
After introduction of local bosonic field operators and spatial
discretization pseudo-Fokker-Planck equations
for multidimensional $s$-parametrized phase-space functions are derived.
These partial differential equations are equivalent to an infinite set of
ordinary differential equations for the cumulants of the phase-space
functions. Introducing an appropriate truncation condition, the resulting
finite set of cumulant evolution equations can be solved numerically.
Solutions are presented in a Gaussian approximation and
the quantum noise is calculated, with special emphasis on
squeezing and the recently measured spectral photon-number
correlations
[Sp\"{a}lter {\it et al}., Phys. Rev. Lett. {\bf 81}, 786 (1998)].
\end{abstract}

%\pacs{PACS number(s): 42.50.Lc,42.50.Dv,42.81.Dp}

% Correlation functions (collective effects), 71.45.G
% Quantum optics, 42.50
% Absorption   -optical, 42.25.B
% Solitons -optical, 42.65.T  ,  -optical fibers, 42.81.D
% Noise -quantum, 42.50.L
% 42.50.Lc Quantum fluctuations, quantum noise, and quantum jumps
% 42.50.Dv Nonclassical field states; squeezed, antibunched, and sub-Poissonian
%          states; operational definitions of the phase of the field; phase
%         measurements
% 42.81.Dp Propagation, scattering, and losses; solitons

\frenchspacing

%\vspace*{-3ex}
\vspace*{2ex}
\narrowtextAAA

\section{Introduction}
\label{Sec.intro}

The study of the nonclassical properties of optical pulses propagating in linear
and nonlinear media has been a subject of increasing interest
\cite{DrummondPD93,SizmannA97,AndersonME97}. In particular effects such as
quadrature-phase squeezing
\cite{DrummondPD93,DrummondPD87,RosenbluhM91,BergmanK94,WernerMJ96},
photon-number squeezing
\cite{WernerMJ96,FribergSR96,SpaelterS97,SchmittS98},
intraspectral quantum correlations \cite{SpaelterS98},
and entanglement \cite{FribergSR92,JoobeurA96}
offer novel possibilities for optical communication and
quantum nondemolition (QND) measurement
\cite{DrummondPD93,SizmannA97}. In this connection the development of
reliable sources of nonclassical light has been of particular interest.
Promising candidates for such sources have been nonlinear optical elements
converting coherent light produced by lasers into nonclassical light.

Since nonclassical properties are always in competition with the noise
associated with unavoidable losses, a quantum theoretical description is
required that necessarily includes absorption. Numerous approaches to the
problem of quantization of radiation in absorbing media have been made (see,
e.g., \cite{SchmidtE97a} and references therein).
In most quantization schemes the influence of absorption has been omitted.
Concepts have been developed for both dispersionless \cite
{JauchJM48,WatsonKM49,PantellEH69,HilleryM84,AbramI87,KnoellL87,MarcuseD89,%
YarivA89,WrightEM90,GlauberRJ91,AbramI91,DeutschIH91,WrightEM91,BlowKJ91,%
BlowKJ92,JoneckisLG93} and (weakly) dispersive media
\cite{DrummondPD87,LaiY89,LaiY89a,DrummondPD90,HuttnerB91,BoivinL94,HilleryM96}.
Formulations in which the medium is introduced macroscopically via a
constant (i.e., frequency-independent) permittivity may run into problems
when dispersion is present \cite{JauchJM48,HilleryM84,GlauberRJ91,AbramI91}.
The problem of quantization of radiation in dispersive and absorbing linear
media has been considered in \cite{AgarwalGS75,FleischhauerM91,HuttnerB92a,%
GrunerT96,KnoellL92,KupiszewskaD92,HoST93,JeffersJ93,GrunerT95,BarnettSM95}.
Recently, the method of quantization via Green-function expansion
\cite{GrunerT96}, which is closely related to the (mesoscopic)
diagonalization scheme considered in \cite{HuttnerB92a}, has been extended
to arbitrary linear media, i.e., arbitrarily space-dependent,
Kramers-Kronig consistent permittivities
\cite{DungHT98a,ScheelS98}. An extension of this approach
to nonlinear, absorbing media has been given in \cite{SchmidtE97a}. An
interesting attempt to treat the problem of quantization of radiation in
dielectric media composed of two-level atoms has been considered in
\cite{HilleryM96}.

In this article the propagation of quantum solitons in absorbing
nonlinear wave guides is studied. It is well known that nonlinear
waveguides are promising tools for ``turn-key''
generation of nonclassical light \cite{SizmannA97},
because of the great nonlinear interaction length and the high
intensity of the radiation \cite{AndersonME97}. Moreover, the radiation
can be well localized in space and time by forming solitonlike pulses.
The stability of classical solitons in a lossless medium is based on the balance
between anomalous group velocity dispersion and
Kerr-induced self-phase modulation
\cite{HasegawaA89,AkhmanovSA92}.
The theoretical description of quantum
solitons usually starts from a Hamiltonian, the classical form of which leads
to the classical (optical) nonlinear Schr\"{o}dinger equation
\cite{WrightEM90,WrightEM91,LaiY89,LaiY89a}.
The quantum nonlinear Schr\"{o}dinger equation (QNLSE) can be solved
by means of the Bethe ansatz \cite{LaiY89a,YaoD95,KartnerFX96}.
In practice, however, the QNLSE must be
supplemented with further terms in order to describe effects such as
absorption, third-order dispersion, and Raman scattering.
Therefore methods
based on the  Hartree approximation
\cite{LaiY89,SingerF92}
and linearization around the classical solution
\cite{HausHA90,LaiY90,LaiY93,DoerrCR94,LaiY95}
have been developed that also apply to the solution of modified QNLSE's.

Alternatively, phase-space methods \cite{GardinerCW91} can be applied \cite
{DrummondPD87,DrummondPD93a,CarterSJ95,WernerMJ97}, starting from the master
equation and introducing --- after appropriate spatial discretization ---
high-dimensional phase-space functions that satisfy
\mbox{(pseu}\-\mbox{do-)} Fok\-ker-Planck equations (PFPE).
The resulting nonlinear, high-dimensional, partial differential equations
are hard to solve in general. In the case of proper FPE's the solution can be
tried to be obtained by stochastic simulation.
In what follows cumulant expansion (see, e.g., \cite{SchackR90,SzlachetkaP93})
is used in order to derive an infinite set of evolution equations for the
cumulant hierarchy associated with a chosen phase-space function. Introducing
appropriate truncation, the resulting finite set of ordinary differential
equations can then be solved numerically. The method is applied
to damped optical quantum solitons in absorbing Kerr media, and
numerical solutions in Gaussian approximation are presented.
The quantum noise is calculated, with special emphasis on
local and spectral
squeezing
\cite{DrummondPD93,DrummondPD87,RosenbluhM91,BergmanK94,WernerMJ96,%
FribergSR96,SpaelterS97,SchmittS98} and
spectral photon-number correlation \cite{SpaelterS98}.

The article is organized as follows. In Sec.~\ref{Sec.master} the underlying
master equation is given. The corresponding pseu\-do-Fok\-ker-Planck
equations for $s$-parametrized phase-space functions are derived in Sec.~\ref
{Sec.FPE}. Section \ref{Sec.cumulant} outlines the cumulant expansion and
the derivation of the cumulant evolution equations. Numerical
results of the
solution in a Gaussian approximation are presented in Sec.~\ref{Sec.results},
and a summary and some concluding remarks are given in Sec~\ref{Sec.summary}.

%%%%%%%%%%%%%%%%%%%%%%%%%%%%%%%%%%%%%%%%%%%%%%%%%%%%%%%%%%%%%%%%%%%%%%%%%%%%%%

\section{Master equation}
\label{Sec.master}

In the simplest model absorption can be described in terms of Markovian
relaxation theory, introducing in the QNLSE a linear damping term and an
additive operator noise source associated with the losses \cite
{LaiY95,CarterSJ95}.
A more rigorous quantization scheme shows that the nonlinear interaction can
also mix the field and the noise source to give a modified QNLSE with
multiplicative noise terms \cite{SchmidtE97a}. Only for sufficiently weak
absorption can the simple model of additive noise be applied. Here we will
restrict our attention to this case.

Let us first consider pulse propagation in a lossless fiber. When the pulse
is long compared to the optical period and spectrally well localized
in an interval $\Delta k$ around the carrier wave number
$k_{\rm c}$,
then the dispersion relation $\omega (k)$ (of the linear wave
equation) can be treated in a second-order approximation,
\begin{equation}
\omega (k_{{\rm c}}+k)\simeq \omega _{{\rm c}}+\omega ^{(1)}k+{\textstyle%
\frac{1}{2}}\omega ^{(2)}k^{2}.
\label{eq.wk}
\end{equation}
Studying the nonlinear media
it may be convenient to define the optical (mid)frequency
$\bar{\omega}_{{\rm c}}$ slightly different from $\omega_{\rm c}$ given in
Eq.~(\ref{eq.wk})
\begin{equation}
\bar{\omega}_{{\rm c}}=\omega _{{\rm c}}-\delta \omega,
\quad \delta \omega \ll \omega_{\rm c}.
\end{equation}
Separating from the (positive- and
negative-frequency parts of the) field the rapidly varying exponentials
$\exp [\mp i(\bar{\omega}_{{\rm c}}t$ $\!-$ $\!k_{{\rm c}}z)]$,
the pulse can be described
in terms of slowly varying annihilation and creation operators
$\hat{b}(z,t)$ and $\hat{b}^{\dagger }(z,t)$, respectively, that satisfy the
equal-time commutation relations \cite{SchmidtE97a}
\begin{eqnarray}
&\left[ \hat{b}(z,t),\hat{b}^{\dagger }(z',t)\right]
 ={\cal A}^{-1}\delta(z-z'),&
\label{eq.aa}
\\
&\left[ \hat{b}(z,t),\hat{b}(z',t)\right]
 =\left[ \hat{b}^{\dagger}(z,t),\hat{b}^{\dagger }(z',t)\right] =0&
\label{eq.aa0}
\end{eqnarray}
(${\cal A}$ is the effective cross section of the fiber).
In what follows it will be convenient to use a reference frame
that moves with the group velocity, i.e.,
\begin{equation}
z = v_{{\rm gr}}t+x,
\label{refsystem}
\end{equation}
where
\begin{equation}
v_{{\rm gr}}=\left. \frac{d\omega }{dk}\right| _{k=k_{c}}=\omega ^{(1)},
\label{group}
\end{equation}
and to introduce the operators
\begin{equation}
\hat{a}(x,t) = \hat{b}(v_{{\rm gr}}t + x ,t),
\quad \hat{a}^\dagger(x,t) = \hat{b}^\dagger(v_{{\rm gr}}t + x ,t).
\label{boperator}
\end{equation}
Obviously, the operators $\hat{a}(x,t)$ and $\hat{a}^{\dagger }(x,t)$
also satisfy commutation relations of the type given in
Eqs.~(\ref{eq.aa}) and (\ref{eq.aa0}) [with
$\delta(x$ $\!-$ $\!x')$ in place of $\delta(z$ $\!-$ $\!z')$].

Under the assumptions made, the undamped spatiotemporal
pulse evolution is governed by the Hamiltonian
\cite{DrummondPD87,WrightEM90,WrightEM91,LaiY89,LaiY89a}
\begin{eqnarray}
\label{hamiltonian}
\lefteqn{ \hat{H}=\hbar {\cal A} \int dx \,
\left[ \vphantomA \delta \omega \hat{a}^{\dagger }\hat{a}\right. }
\nonumber \\
&&\quad\quad\quad+\left. {\textstyle\frac{1}{2}} \omega^{(2)} \left(
\partial _{x}\hat{a}^{\dagger }\right) \left( \partial _{x}\hat{a}\right)
+ {\textstyle\frac{1}{2}}\chi
\hat{a}^{\dagger }\hat{a}^{\dagger }\hat{a}\hat{a}\vphantomA\right] ,
\end{eqnarray}
where the constant ${\chi }$ is related to the third-order susceptibility
$\chi^{(3)}$ as \cite{DrummondPD87}
\begin{equation}
{\chi }=\frac{3\chi ^{(3)}\hbar (v_{{\rm gr}}k_{{\rm c}})^{2}}
{4\epsilon_{{\rm r}}^{2}\epsilon _{0}}
\end{equation}
($\epsilon_{{\rm r}}$ is the relative permittivity at frequency
$\omega_{\rm c}$).
Note that solitons can be formed either
in focusing media with anomalous dispersion
(\mbox{$\chi$ $\!<$ $0$}, \mbox{$\omega^{(2)}$ $\!>$ $\!0$}) or
in defocusing media with normal dispersion
(\mbox{$\chi$ $\!>$ $\!0$}, \mbox{$\omega^{(2)}$ $\!<$ $\!0$}).

Including in the theory damping as outlined above, the density operator
(in the Schr\"odinger picture) satisfies the master equation
\begin{equation}
i\hbar ~\partial _{t}\hat{\rho}=[\hat{H},\hat{\rho}]
+i\gamma \hat{L}\hat{\rho},
\label{eq.master}
\end{equation}
where the Lindblad term \cite{LindbladG76}
\begin{eqnarray}
\lefteqn{ \gamma \hat{L}\hat{\rho}
=\gamma \hbar {\cal A}\int dx \,
\left[\vphantomA N_{\rm th}\left( 2\hat{a}^{\dagger }\hat{\rho}\hat{a}
-\hat{\rho}\hat{a}\hat{a}^{\dagger }
-\hat{a}\hat{a}^{\dagger }\hat{\rho}\right) \right. }
\nonumber \\
&&\quad\quad\quad + \left.
\left( N_{\rm th}+1\right) \left( 2\hat{a}\hat{\rho}\hat{a}^{\dagger }
-\hat{\rho}\hat{a}^{\dagger }\hat{a}
-\hat{a}^{\dagger }\hat{a}\hat{\rho}\right) \vphantomA\right]
\end{eqnarray}
is responsible for damping, with $\gamma $ being the damping constant, and
\begin{equation}  \label{Nthermisch}
N_{\rm th}=\left[ \exp \left( \frac{\omega_{{\rm c}}\hbar }{k_{{\rm B}}T}\right)
-1\right] ^{-1}
\end{equation}
($k_{{\rm B}}$  is the Boltzmann constant, $T$  the absolute temperature).

To obtain a discrete version of Eq.~(\ref{eq.master}), we divide the
$x$ interval
into $m$ cells of size $\Delta x$ numbered with
index $j$ and introduce discrete operators
\cite{CarterSJ95}
\begin{equation}
\hat{a}_{j}(t)=\sqrt{\frac{{\cal A}}{\Delta x}}\int_{x_{j}}^{x_{j}+\Delta x}dx\,
\hat{a}(x,t),
\label{aj}
\end{equation}
with
\begin{equation}
x_{j}=\Delta x (j-[m/2]).
\label{xj}
\end{equation}
It follows from Eqs.~(\ref{eq.aa}) and (\ref{eq.aa0}) that the
operators $\hat{a}_{j}$ and $\hat{a}_{j}^{\dagger }$ satisfy the equal-time
commutation relations
\begin{eqnarray}
&\left[ \hat{a}_{j}(t),\hat{a}_{j'}^{\dagger }(t)\right]
=\delta_{jj'},&
\label{commutatoraj1}
\\
&\left[ \hat{a}_{j}(t),\hat{a}_{j'}(t)\right] =
\left[ \hat{a}_{j}^{\dagger }(t),\hat{a}_{j'}^{\dagger }(t)\right] =0.&
\label{commutatoraj2}
\end{eqnarray}
When $\Delta x$ is sufficiently small, then Eq.~(\ref{aj}) reduces to
\begin{equation}
\hat{a}_{j}(t)\simeq \hat{a}(x_{j},t)\sqrt{\Delta x {\cal A}},
\label{ersetzung1}
\end{equation}
and the relations for continuous quantities can be replaced with the
relations for discrete quantities
according to the rules
\begin{eqnarray}
\lefteqn{
\int dx \, {\cal F}\!\left[\hat{a}(x),\hat{a}^\dagger(x)\right]
}
\nonumber\\&&\hspace{4ex}
\simeq
\sum_{j} \Delta x \, {\cal F}\!\left[\hat{a}_j/\sqrt{\Delta x {\cal A}},
\hat{a}_j^\dagger/\sqrt{\Delta x {\cal A}}\right] ,
\label{ersetzung2}
\\&&
\partial _{x}\hat{a}
\simeq \frac{1}{\sqrt{\Delta x {\cal A}}}
\frac{\hat{a}_{j+1}-\hat{a}_{j}}{\Delta x}\,,
\label{ersetzung3}
\end{eqnarray}
where ${\cal F}[\hat{a}(x),\hat{a}^\dagger(x)]$ is an arbitrary function
of $\hat{a}(x)$ and $\hat{a}^\dagger(x)$.
Finally, it is often convenient to describe pulse propagation
in terms of appropriately scaled quantities. In our
numerical calculations we have used the scaled quantities given
in Appendix~\ref{Sec.scaled.vars}.

%%%%%%%%%%%%%%%%%%%%%%%%%%%%%%%%%%%%%%%%%%%%%%%%%%%%%%%%%%%%%%%%%%%%%%%%%%%%%

\section{Fokker-Planck equations}
\label{Sec.FPE}

It is well known that the description of the quantum state in terms of the
density matrix is fully equivalent to a description in terms of phase-space
functions. In what follows we restrict our attention to $s$-parametrized
phase-space functions \cite{CahillKE69a,CahillKE69b}, which for $m$
operators $\hat{a}_{j}$ are  functions of $m$ complex phase-space variables
${a}_{j}$. For notational convenience we will omit the time argument
and write the phase-space arguments in a
compact vector form
\begin{equation}
P(a_{1},a_{2},\ldots,a_{m},t;s)\equiv P({\bf a};s)=P({\bf u},{\bf v};s),
\label{psfunction}
\end{equation}
with the real vectors ${\bf u}$ and ${\bf v}$ being defined by
\begin{equation}
{\bf a}={\bf u}+i{\bf v}.
\label{eq.a.uv}
\end{equation}
Note that expectation values of $s$-ordered operator pro\-ducts
are obtained by averaging the corresponding pro\-ducts of
complex phase-space variables with the $s$-parametrized phase-space
function, i.e.,
\begin{eqnarray}
\lefteqn{
\langle \hat{a}_1^{\dagger}{^{n_1}} \dots \hat{a}_m^{\dagger}{^{\!\!\!n_m}}
\hat{a}_1^{k_1} \dots \hat{a}_m^{k_m} \rangle_s
}
\nonumber\\ && \hspace{2ex}
  = \int d^m{\bf u}\, d^m{\bf v} \, \big[
  \left(u_1-iv_1\right)^{n_1} \dots
  \left(u_m-iv_m\right)^{n_m}
\nonumber\\ && \hspace{2ex} \times
  \left(u_1+iv_1\right)^{k_1} \dots
  \left(u_m+iv_m\right)^{k_m}
  \big]  P({\bf u},{\bf v};s) ,
\label{eq.aa.uv}
\end{eqnarray}
where the notation $\langle\hat{a}_1^{\dagger n_1}\dots
\hat{a}_m^{k_m}\rangle_s$ is used to introduce $s$ ordering.

Applying standard rules (see, e.g., \cite{VogelK89}), the evolution equation for
$P({\bf u},{\bf v};s)$ can be derived from the master equation
straightforwardly. After some algebra we obtain
\widetextEQ
\begin{eqnarray}
\lefteqn{
\partial _{t}P({\bf u},{\bf v};s)
=\sum_{j}\left\{ \vphantomAAA-\delta\omega
\left[ \frac{\partial }{\partial u_{j}}v_{j}-\frac{\partial }
{\partial v_{j}}u_{j}\right] \right. -\frac{\omega ^{(2)}}{2\Delta x^{2}}
\left[ \frac{\partial }{\partial u_{j}}(2v_{j}\!-\!v_{j+1}\!-\!v_{j-1})
-\frac{\partial }{\partial v_{j}}(2u_{j}\!-\!u_{j+1}\!-\!u_{j-1})\right]
}
\nonumber \\ && \hspace{2ex}
+\,\bar{\chi}\left[ \frac{\partial }{\partial u_{j}}v_{j}-\frac{\partial}
{\partial v_{j}}u_{j}\right] \left( 1\!-\!s\!-\!|a_{j}|^{2}\right)
+\gamma \left[ \frac{\partial }{\partial u_{j}}u_{j}+\frac{\partial }
{\partial v_{j}}v_{j}
\right]
+{\bar{\chi}}\frac{1\!-\!s^{2}}{16}
\left( \frac{\partial ^{2}}{\partial u_{j}^{2}}+\frac{\partial ^{2}}
{\partial v_{j}^{2}}\right) \left( \frac{\partial }{\partial u_{j}}v_{j}
-\frac{\partial }{\partial v_{j}}u_{j}\right) 
\nonumber \\ && \hspace{2ex}
\left. 
+\,{\bar{\chi}}\frac{s}{2}\left[ \left( \frac{\partial
^{2}}{\partial u_{j}^{2}}-\frac{\partial ^{2}}{\partial v_{j}^{2}}\right)
u_{j}v_{j}-\frac{\partial }{\partial u_{j}}\frac{\partial }{\partial v_{j}}
\left( u_{j}^{2}-v_{j}^{2}\right) \right] 
+\frac{\gamma }{2}
\left( N_{\rm th}+\frac{1\!-\!s}{2}\right)
\left( \frac{\partial
^{2}}{\partial u_{j}^{2}}+\frac{\partial ^{2}}{\partial v_{j}^{2}}\right)
\vphantomAAA \right\}
P({\bf u},{\bf v};s)
\label{eq.FPE.uv}
\end{eqnarray}
\narrowtextEQ
\noindent
[$\bar{\chi}$ $\!=$ $\!\chi/(\Delta x {\cal A})$].
An equation of this type is also called the pseudo-Fokker-Planck equation.
Apart from the fact that in an ordinary FPE derivatives only up to
second order appear \cite{RiskenH84}, it is impossible, in general, to find
a stochastic process that is equivalent to Eq.~(\ref{eq.FPE.uv}) and can
be used for solving the equation.
For $s$ $\!=$ $\!1$ ($P$ function) and $s$ $\!=$ $\!-1$ ($Q$ function) 
the two last terms in curly brackets
in Eq.~(\ref{eq.FPE.uv}) correspond to
noise with a nonpositive diffusion matrix which cannot be modeled by a
stochastic process \cite{DrummondPD87}. Nevertheless, such a PFPE can
have nonsingular solutions (see, e.g., \cite{VogelK88}).
For $s$ $\!=$ $\!0$ (Wigner function)
the fifth term in curly brackets in Eq.~(\ref{eq.FPE.uv}) contains so
called third-order noise which also cannot be treated by
stochastic simulation \cite{GardinerCW85}.
For all other values of $s$ the two terms appear
together. It should be pointed out that for the so-called positive $P$
representation the master equation (\ref{eq.master}) corresponds to a proper
FPE with positive definite diffusion matrix \cite
{DrummondPD87,CarterSJ95,WernerMJ97b}, so that stochastic simulation applies
(see, e.g., \cite{WernerMJ97a}). However, the price to be paid is the
doubling of variables.

%%%%%%%%%%%%%%%%%%%%%%%%%%%%%%%%%%%%%%%%%%%%%%%%%%%%%%%%%%%%%%%%%%%%%%%%%%%%%%
%%%%%%%%%%%%%%%%%%%%%%%%%%%%%%%%%%%%%%%%%%%%%%%%%%%%%%%%%%%%%%%%%%%%%%%%%%%%%%

\section{Cumulant expansion}
\label{Sec.cumulant}

In order to solve Eq.~(\ref{eq.FPE.uv}), we apply cumulant expansion
techniques. For this purpose we express the characteristic function
of $P({\bf u},{\bf v};s)$ in terms of cumulants and derive
the cumulant evolution equations. After appropriate truncation
the set of ordinary differential equations can be solved numerically.

%%%%%%%%%%%%%%%%%%%%%%%%%%%%%%%%%%%%%%%%%%%%%%%%%%%%%%%%%%%%%%%%%%%%%%%%%%%%%%

\subsection{Basic definitions}

Let us consider a (quasi)probability distribution function
$P({\bf q})$ $\!\equiv$ $P(q_1,q_2,\ldots,q_f)$ of $f$ real variables
$q_k$ and define its characteristic function
$\chi(\underline{\bf q})$ $\!\equiv$ $\!\chi(\underline{q}_1,
\underline{q}_2,\ldots,\underline{q}_m)$
by the Fourier transform of $P({\bf q})$,
\begin{eqnarray}
P({\bf q}) =\frac{1}{(2\pi )^{f}}\int d^{f}\underline{\bf q}\,
\chi (\underline{\bf q})e^{-i\underline{\bf q}\cdot{\bf q}},
\label{eq.distr.func}
\end{eqnarray}
\begin{eqnarray}
\chi(\underline{\bf q}) =\int d^{f}{\bf q}\,
P({\bf q})e^{i\underline{\bf q}\cdot{\bf q}}.
\label{eq.char.func}
\end{eqnarray}
   From Eq.~(\ref{eq.char.func}) it follows that the moments
$\langle q_{1}^{n_{1}}\ldots q_{f}^{n_{f}}\rangle$
of the distribution function $P({\bf q})$ can be obtained from
$\chi(\underline{\bf q})$ by differentiation,
\begin{eqnarray}
\lefteqn{
\langle q_1^{n_{1}}\ldots q_f^{n_{f}}\rangle
= \int d^{f}{\bf q}\,
q_{1}^{n_{1}}\ldots q_f^{n_{f}}P({\bf q})
}
\nonumber\\ && \hspace{6ex}
= \left. \left( \frac{\partial }{i\partial \underline{q}_{1}}\right) ^{n_{1}}\ldots
\left( \frac{\partial }{i\partial \underline{q}_{f}}\right) ^{n_{f}}\chi (\underline{\bf q}
)\right| _{\underline{\bf q}=0}
\end{eqnarray}
($n_k$ $\!\ge$ $\!0$).
Hence, the Taylor-series expansion of the function $\chi(\underline{\bf q})$,
which can be regarded as moment generating function, reads as
\begin{eqnarray}
\chi (\underline{\bf q}) = \sum_{\{n_k\}}
\langle q_1^{n_{1}}\ldots q_f^{n_{f}}\rangle \,
\frac{(i\underline{q}_{1})^{n_{1}}\ldots (i\underline{q}_{f})^{n_{f}}}
{n_{1}!\ldots n_{f}!} \,.
\label{eq.mean.series}
\end{eqnarray}
Introducing the cumulant generating function $\Phi (\underline{\bf q})$,
\begin{eqnarray}
\chi (\underline{\bf q}) = e^{\Phi (\underline{\bf q})} ,
\label{eq.chi.phi}
\end{eqnarray}
the cumulants $\langle\! \langle q_{1}^{n_{1}}\ldots
q_{f}^{n_{f}}\rangle\!\rangle$ are defined by the
Taylor-series expansion of $\Phi (\underline{\bf q})$ as follows:
\begin{eqnarray}
\Phi (\underline{\bf q}) = \sum_{\{n_k\}} \langle
\!\langle q_{1}^{n_{1}}\ldots q_{f}^{n_{f}}\rangle\!\rangle
\,\frac{(i\underline{q}_{1})^{n_{1}}\ldots (i\underline{q}_{f})^{n_{f}}}{n_{1}!\ldots n_{f}!}\,.
\label{eq.cums.series}
\end{eqnarray}
Combining Eqs.~(\ref{eq.mean.series}) -- (\ref{eq.cums.series}),
the cumulants and moments can be related to each other as
\begin{eqnarray}
\lefteqn{
\langle q_1^{n_{1}}\ldots q_f^{n_{f}}\rangle
=\left( \frac{\partial}{\partial \underline{q}_{1}}\right)^{n_{1}}\ldots
\left( \frac{\partial }{\partial \underline{q}_{f}}\right)^{n_{f}}
}
\nonumber\\[1ex]&&\hspace{2ex}
\label{eq.means}
\times \left. \exp \!\left[ \sum_{\{k_l\}}
\langle\!\langle q_{1}^{k_{1}}\ldots q_{f}^{k_{f}}\rangle\!\rangle
\,\frac{\underline{q}_{1}^{k_1}\ldots \underline{q}_{f}^{k_{f}}}{k_{1}!\ldots k_{f}!}
\right] \right|_{\underline{\bf q}=0},
\end{eqnarray}
\begin{eqnarray}
\lefteqn{
\langle\!\langle q_1^{n_{1}}\ldots q_f^{n_{f}}\rangle\!\rangle
=\left( \frac{\partial }{\partial \underline{q}_{1}}\right)^{n_{1}}\ldots
\left( \frac{\partial }{\partial \underline{q}_{f}}\right)^{n_{f}}
}
\nonumber\\[1ex]&&\hspace{2ex}
\label{eq.cums}
\times \left. \ln\!\left[ \sum_{\{k_l\}}
\langle q_1^{k_{1}}\ldots q_f^{k_{f}}\rangle
\,\frac{\underline{q}_{1}^{k_1}\ldots \underline{q}_{f}^{k_{f}}}{k_{1}!\ldots k_{f}!}
\right] \right|_{\underline{\bf q} =0}
\end{eqnarray}
($\langle\!\langle q_1^{0}\ldots q_f^{0}\rangle\!\rangle$ $\!=$ $ \!0$).
Note that the symmetry relations
\begin{eqnarray}
\langle \ldots q_{i}^{n_{i}}\ldots q_{j}^{n_{j}}\ldots \rangle
=\langle \ldots q_{j}^{n_{j}}\ldots q_{i}^{n_{i}}\ldots \rangle,
\label{eq.sym.md}
\end{eqnarray}
\begin{eqnarray}
\langle\!\langle \ldots q_{i}^{n_{i}}\ldots q_{j}^{n_{j}}\ldots
\rangle\!\rangle
= \langle \!\langle \ldots
q_{j}^{n_{j}}\ldots q_{i}^{n_{i}}\ldots \rangle \!\rangle
\label{eq.symm.cd}
\end{eqnarray}
are valid. For a Gaussian distribution higher than second-order cumulants
are equal to  zero,
\begin{eqnarray}
\langle \!\langle q_{1}^{n_{1}}\ldots q_{f}^{n_{f}}\rangle \!\rangle = 0
\quad \text{if} \quad n_{1}+\ldots +n_{f} > 2,
\end{eqnarray}
so that higher than second-order moments
can be expressed in terms of first- and second-order cumulants
$\langle\!\langle q_{i}\rangle\!\rangle $
and $\langle\!\langle q_{i}q_{j}\rangle\!\rangle $,
respectively (see, e.g., \cite{GardinerCW85}).

Obviously, Eqs.~(\ref{eq.distr.func}) - (\ref{eq.symm.cd})
also apply to the $s$-para\-metrized phase-space functions
$P({\bf u},{\bf v};s)$ introduced in Sec.~\ref{Sec.FPE}
[$P({\bf q})$ $\!\to$ $\!P({\bf u},{\bf v};s)$,
$\chi(\underline{\bf q})$ $\!\to$ $\!\chi(\underline{\bf u},\underline{\bf v};s)$,
$\Phi(\underline{\bf q})$ $\!\to$ $\!\Phi(\underline{\bf u},\underline{\bf v};s)$].
Since the characteristic functions for different values of $s$
are related to each other as \cite{CahillKE69b}
\begin{eqnarray}
\lefteqn{
\chi (\underline{\bf u},\underline{\bf v};s)
= \chi (\underline{\bf u},\underline{\bf v};s')
}
\nonumber\\[1ex] && \hspace{2ex}
\times \, \exp\!\left[ {\textstyle\frac{1}{8}}(s-s')
\left(\underline{u}_{1}^{2}+\underline{v}_{1}^{2}+\ldots
+ \underline{u}_{m}^{2}+\underline{v}_{m}^{2}\right) \right] ,
\end{eqnarray}
from Eqs.~(\ref{eq.chi.phi}) and (\ref{eq.cums.series}) it is
seen that only the second-order cumulants of equal variables
depend on $s$,
\begin{eqnarray}
\langle\!\langle u_{j}u_{j'}\rangle\!\rangle_{s}
&=& \langle\!\langle u_{j}u_{j'}\rangle\!\rangle_{s'}
-{\textstyle\frac{1}{4}}(s-s')\delta_{jj'}
\label{eq.C20.s}
\end{eqnarray}
and $\langle\!\langle v_{j}v_{j'}\rangle\!\rangle_{s}$ accordingly.
All other cumulants are independent of $s$ \cite{SchackR90}.
Note that the mean value of the complex amplitude
$\langle\hat{a}_j\rangle$ and the intensity
$I_j$ $\!\equiv$ $\!\langle\hat{a}^\dagger_j\hat{a}_j\rangle$ are given by
\begin{eqnarray}
&\langle\hat{a}_j\rangle =
\langle\!\langle u_j \rangle\!\rangle_s
+ i \langle\!\langle v_j \rangle\!\rangle_s,&
\label{eq.mean.aj}
\\[1ex]
&I_j =
\langle\!\langle u_j \rangle\!\rangle_s^2
+\langle\!\langle v_j \rangle\!\rangle_s^2
+\langle\!\langle u_j^2 \rangle\!\rangle_s
+\langle\!\langle v_j^2 \rangle\!\rangle_s
+{\textstyle\frac{1}{2}}(s-1) \,.&
\label{eq.Ij}
\end{eqnarray}

%%%%%%%%%%%%%%%%%%%%%%%%%%%%%%%%%%%%%%%%%%%%%%%%%%%%%%%%%%%%%%%%%%%%%%%%%%%%%%

\subsection{Evolution equations}

Let us assume that a (quasi)probability distribution function
$P({\bf q})$ satisfies an evolution equation of the following type:
\begin{eqnarray}
\lefteqn{
\partial _{t}P({\bf q})
= \sum_{\{n_l\},\{k_l\}} \bigg[ A_{n_1\ldots n_f}^{k_1\ldots k_f}
}
\nonumber\\&&\hspace{4ex}\times
\left( \frac{\partial }{\partial q_{1}}\right)^{n_{1}}q_{1}^{k_{1}}\ldots
\left( \frac{\partial }{\partial q_{f}}\right)^{n_{f}}q_{f}^{k_{f}}
\bigg] P({\bf q}) ,
\label{eq.FPE.one}
\end{eqnarray}
where $A_{n_1\ldots n_f}^{k_1\ldots k_f}$ are constants.
It follows from Eq.~(\ref{eq.FPE.one})
that the characteristic function $\chi (\underline{\bf q})$,
Eq.~(\ref{eq.char.func}), satisfies the evolution equation
\begin{eqnarray}
\lefteqn{
\partial_{t}\chi (\underline{\bf q}) = \partial_{t}
\int d^{f}{\bf q} \, e^{i\underline{\bf q}\cdot{\bf q}} P({\bf q})
}
\nonumber\\&&\hspace{2ex}
= \sum_{\{n_l\},\{k_l\}} \bigg[ A_{n_1\ldots n_f}^{k_1\ldots k_f}
(-i\underline{q}_{1})^{n_{1}}\ldots (-i\underline{q}_{f})^{n_{f}}
\nonumber\\&&\hspace{4ex}\times
\left( \frac{\partial }{i\partial \underline{q}_{1}}\right)^{k_{1}}
\ldots \left( \frac{\partial }{i\partial \underline{q}_{f}}\right)^{k_{f}}
\bigg]\chi (\underline{\bf q}),
\end{eqnarray}
and hence [see Eq.~(\ref{eq.chi.phi})]
\begin{eqnarray}
\lefteqn{
\partial_{t}\Phi (\underline{\bf q}) =
e^{-\Phi (\underline{\bf q})}
\!\!\!\!
\sum_{\{n_l\},\{k_l\}}\!\! \bigg[ A_{n_1\ldots n_f}^{k_1\ldots k_f}
(-i\underline{q}_{1})^{n_{1}}\!\ldots (-i\underline{q}_{f})^{n_{f}}
}
\nonumber\\&&\hspace{12ex}\times
\left( \frac{\partial }{i\partial \underline{q}_{1}}\right)^{k_{1}}
\ldots \left( \frac{\partial}{i\partial \underline{q}_{f}}\right)^{k_{f}}
e^{\Phi (\underline{\bf q})}
\bigg].
\label{eq.dC.dt}
\end{eqnarray}
Substituting for $\Phi(\underline{\bf q})$ in Eq.~(\ref{eq.dC.dt}) the
Taylor-series expansion (\ref{eq.cums.series})
and comparing the coefficients of equal powers we arrive at
\begin{eqnarray}
\label{eq.evol.cums}
\lefteqn{
\partial _{t}\langle\!\langle q_{1}^{r_{1}}\ldots
q_{f}^{r_{f}}\rangle\!\rangle
= \sum_{\{n_l\},\{k_l\}}   \bigg[
A_{n_1\ldots n_f}^{k_1\ldots k_f}
}
\nonumber\\&&\hspace{4ex}
\times \, \frac{(-i)^{n_{1}+k_{1}+r_{1}}r_{1}!}{(r_{1}-n_{1})!}%
\ldots
\frac{(-i)^{n_{f}+k_{f}+r_{f}}r_{f}!}{(r_{f}-n_{f})!}
\nonumber \\&&\hspace{4ex}
\times \left( \frac{\partial }{\partial \underline{q}_{1}}\right)^{r_{1}-n_{1}}
\ldots
\left( \frac{\partial }{\partial \underline{q}_{f}}\right)^{r_{f}-n_{f}}
e^{-\Phi (\underline{\bf q})}
\nonumber \\&&\hspace{4ex}
\times  \left( \frac{\partial }{\partial \underline{q}_{1}}\right)^{k_{1}}
\ldots
\left.
\left( \frac{\partial }{\partial \underline{q}_{f}}\right) ^{k_{f}}
e^{\Phi (\underline{\bf q})} \bigg] \right| _{\underline{\bf q}=0}
\end{eqnarray}
($1/n!$ $\!\equiv$ $\!0$ if $n$ $\!<$ $\!0$).
Performing in Eq.~(\ref{eq.evol.cums}) the differentiations
yields the cumulant evolution equations explicitly.

We now apply the scheme to the PFPE (\ref{eq.FPE.uv}) in order
to obtain the evolution equations of the cumulants associated
with $P({\bf u},{\bf v};s)$. For the sake of clearness we
somewhat change the notation, using
$C_{u^n\cdots,v^k\cdots,\ldots}(x,x',\ldots)$ to denote
the cumulants (for chosen $s$), e.g.,
\begin{equation}
C_{uv,u^2}(x,x')
\equiv \langle\!\langle u_{j}v_{j}u_{j'}^{2} \rangle\!\rangle _{s}
\quad (j\neq j') ,
\label{eq.Cxy.def}
\end{equation}
with
$x$ $\!\equiv x_{j}$
[cf. Eq.~(\ref{xj}); note that $x_{j\pm 1}$ $\!=$ $\!x \pm \Delta x$].
After some algebra the evolution equations for the first-order cumulants
$C_{u}(x)$ and $C_{v}(x)$ are derived to be
\begin{eqnarray}
\label{eq.Cu}
\lefteqn{
\partial _{t}C_{u}(x)=-\gamma \,C_{u}(x)+\delta \omega \,C_{v}(x)
}
\nonumber\\&&\hspace{2ex}
+\,\frac{\omega ^{(2)}}{2\Delta x^{2}}{\,\left[
2\,C_{v}(x)-C_{v}(x\!-\!\Delta x)-C_{v}(x\!+\!\Delta x)\right] }
\nonumber\\&&\hspace{2ex}
+\,{\bar{\chi}}\,C_{v}(x)\left[ {{C_{u}^{2}(x)}}\,+{{C_{v}^{2}(x)}}
\right]
\nonumber\\&&\hspace{2ex}
+\,{\bar{\chi}}\,C_{v}(x)\left[
(s-1)+C_{u^2}(x)+3C_{v^2}(x)\right]
\nonumber\\&&\hspace{2ex}
+\,2{\bar{\chi}}C_{u}(x)\,C_{uv}(x)
+{\bar{\chi}}\left[ C_{u^2v}(x)+C_{v^3}(x)\right] ,
\end{eqnarray}
\begin{eqnarray}
\label{eq.Cv}
\lefteqn{
\partial _{t}C_{v}(x)=-\gamma \,C_{v}(x)-\delta \omega \,C_{u}(x)
}
\nonumber\\&&\hspace{2ex}
+\,\frac{\omega ^{(2)}}{2\Delta x^{2}}{\,\left[
-2\,C_{u}(x)+C_{u}(x\!-\!\Delta x)+C_{u}({x}\!+\!\Delta x)\right] }
\nonumber\\&&\hspace{2ex}
-\,{\bar{\chi}}\,{{C_{u}(x)}}\left[ {{C_{u}^{2}(x)}}\,+{{C_{v}^{2}(x)}}
\right]
\nonumber\\&&\hspace{2ex}
-\,{\bar{\chi}}\,{{C_{u}(x)}}\left[
(s-1)+3C_{u^2}(x)+C_{v^2}(x)\right]
\nonumber\\&&\hspace{2ex}
-\,2\,{\bar{\chi}}\,C_{v}(x)\,C_{uv}(x)
-\,{\bar{\chi}}\left[ C_{u^3}(x)+C_{uv^2}(x)\right] .
\end{eqnarray}
Owing to the nonlinear interaction the first-order cumulants
are coupled to second- and third-order cumulants, whose
evolution equations are rather lengthy
[for the evolution equations for the second-order cumulants,
see Eqs.~(\ref{eq.Cuu}) - (\ref{eq.Cvv}) in Appendix~\ref{Sec.2nd.cums}].
Note that the right-hand sides of Eqs.~(\ref{eq.Cu}), (\ref{eq.Cv}),
and (\ref{eq.Cuv})
are $s$ independent [in accordance with Eq.~(\ref{eq.C20.s}) and the
statement given there].
Note that in the limiting case
$\Delta x$ $\!\rightarrow$ $\!0$
the cumulant evolution equations
[cf. Eqs.~(\ref{eq.Cu}), (\ref{eq.Cv}), 
 and (\ref{eq.Cuu})-(\ref{eq.Cvv})]
are transformed into partial differential equations
for corresponding cumulant correlation functions.

%%%%%%%%%%%%%%%%%%%%%%%%%%%%%%%%%%%%%%%%%%%%%%%%%%%%%%%%%%%%%%%%%%%%%%%%%%%%%%

\subsection{Gaussian approximation}

With regard to a numerical treatment, the hierarchy of cumulant
evolution equations, which represents an infinite set of coupled,
nonlinear, ordinary, first-order differential equations,
must be decomposed into finite, closed systems
of equations. For this purpose approximation schemes
are required to be applied.
Note that only for the linear
system (${\bar{\chi}}$ $\!=$ $\!0$) does the infinite set of
equations decompose into finite, closed subsets, each of which
contains cumulants of the same order.

For a sufficiently  high
peak photon number of the order of magnitude of
$\bar{n}$ $\!\sim$ $\!10^{9}$
[Eq.~(\ref{nbar})] an expansion in $1/\bar{n}$
may be performed,
in which the second-order cumulants are considered to be
of  magnitude  $1/\bar{n}$ in comparison with the first-order cumulants
and cumulants of higher than second order are neglected.
Comparing terms of equal order we
obtain a linear system of differential equations
for the second-order cumulants, in which
the first-order cumulants
governed by truncated Eqs.~(\ref{eq.Cu}) and (\ref{eq.Cv})
play the role of time-dependent
coefficients (for such linearization schemes for lossless systems, see
\cite{DrummondPD87,HausHA90,LaiY90,LaiY93,DoerrCR94},
and for absorbing case, e.g., \cite{LaiY95}).
The approximation scheme may be extended straightforwardly
in order to also include higher-order cumulants in the
calculation.

Another approach to the problem is to retain all cumulants up
to some appropriately chosen order ${\cal N}$ and to set the remaining
higher-order cumulants equal to zero.
In what follows we restrict our attention to the
second-order (Gaussian) approximation (${\cal N}$ $\!=$ $\!2$).
Thus in Eqs.~(\ref{eq.Cu}), (\ref{eq.Cv}), and (\ref{eq.Cuu}) --
(\ref{eq.Cvv}) all terms that contain cumulants of third and fourth
order are omitted, and the now closed system of
nonlinear differential equations is solved numerically.
This approximation obviously corresponds to the assumption
that the quantum state of the pulse can be approximated by a
Gaussian phase-space function $P({\bf u},{\bf v};s)$.
Note that for Gaussian distributions
higher than second order cumulants exactly vanish.

The Gaussian approximation may be expected to apply if the damping
is not too small. To roughly estimate the range of validity of the Gaussian
approximation, we note that the linearization approach used in
\cite {HausHA90} for an undamped
fundamental
quantum soliton fails
at (scaled) times $\tilde{t}$ $\!\sim$  $\!\bar{n}^{1/4}$
\cite{KartnerFX96}
(for scaled quantities, see Appendix~\ref{Sec.scaled.vars}).
On the other hand, in absorbing media the soliton noise
relaxes to thermal-equilibrium noise
on a time scale $\sim$ $\!\tilde{\gamma}^{-1}$.
As the non-Gaussian correlations are suppressed more rapidly than
the Gaussian ones it is expected that
for $\tilde{\gamma}^{-1}$ $\!\le$  $\!\bar{n}^{1/4}$
the Gaussian approximation is valid for all times
of pulse propagation.
Hence, the lower limit for the scaled absorption coefficient is given by
\begin{equation}
\tilde{\gamma}
\gtsim
\bar{n}^{-1/4}
\label{eq.gam.min},
\end{equation}
which for $\bar{n}$ $\!\sim$ $\!10^{9}$ yields
$\tilde\gamma$ $\!\gtsim$ $\!5.6\times 10^{-3}$.
For smaller values of $\tilde\gamma$ the Gaussian approximation
is expected to be valid only in a limited time interval
$\tilde{t}$ $\!\ltsim$ $\!\bar{n}^{1/4}$.

%%%%%%%%%%%%%%%%%%%%%%%%%%%%%%%%%%%%%%%%%%%%%%%%%%%%%%%%%%%%%%%%%%%%%%%%%%%%%
%%%%%%%%%%%%%%%%%%%%%%%%%%%%%%%%%%%%%%%%%%%%%%%%%%%%%%%%%%%%%%%%%%%%%%%%%%%%%%

\section{Results}
\label{Sec.results}

In the Gaussian approximation
the steady-state solution of
Eqs.~(\ref{eq.Cu}), (\ref{eq.Cv}), and (\ref{eq.Cuu}) --
(\ref{eq.Cvv}) can be found to be
\begin{eqnarray}
&{C}_{u}(x) = {C}_{v}(x) = 0 ,&
\label{eq.C1.stat}\\
&{C}_{u,u}(x,x')\! =\! {C}_{v,v}(x,x')
\!= \!\textstyle{\frac{1}{2}}\!
\left[N_{\rm th}\!+\!\textstyle{\frac{1}{2}}(1\!-\!s)\right] \!\delta(x,x'),&
\label{eq.Cuu.stat}\\
&{C}_{u,v}(x,x') = {C}_{v,u}(x,x') = 0&
\label{eq.Cuv.stat}
\end{eqnarray}
[$\delta(x,x')$ $\!=$ $\!1$, if $|x$ $\!-$ $\!x'|$ $\!\le$ $\!\Delta x/2$,
and  $\delta(x,x')$ $\!=$ $\!0$ otherwise].
It corresponds to thermal fluctuations, without any
correlation between the fields at different space points. In the
numerical solution of the time-dependent problem we have used the
steady-state values (\ref{eq.Cuu.stat}) and (\ref{eq.Cuv.stat})
as initial values of the second-order cumulants. With regard to
the initial values of the first-order cumulants, we have used
the fundamental soliton solution of the classical nonlinear
Schr\"{o}dinger equation \cite{HasegawaA89,AkhmanovSA92},
\begin{eqnarray}
{C}_{u}(x) &=& \sqrt{n_0} \,
\mathop{\rm sech}(x/x_{0}),
\label{eq.Cu.ini}\\
{C}_{v}(x) &=& 0
\label{eq.Cv.ini}
\end{eqnarray}
\if\EprintServer\YES
%\def\figINT{y}
%\input{figures.inc}
%\if\figINT\YES % int
\begin{minipage}{\widthAAA}
\begin{figure}
\par
  \if\Figures\YES
%     \hbox{\centerline{\psfig{figure=int.ps,height=\heightFIG,bbllx=75pt,bblly=80pt,bburx=340pt,bbury=360pt}}}
    \hbox{\centerline{\psfig{figure=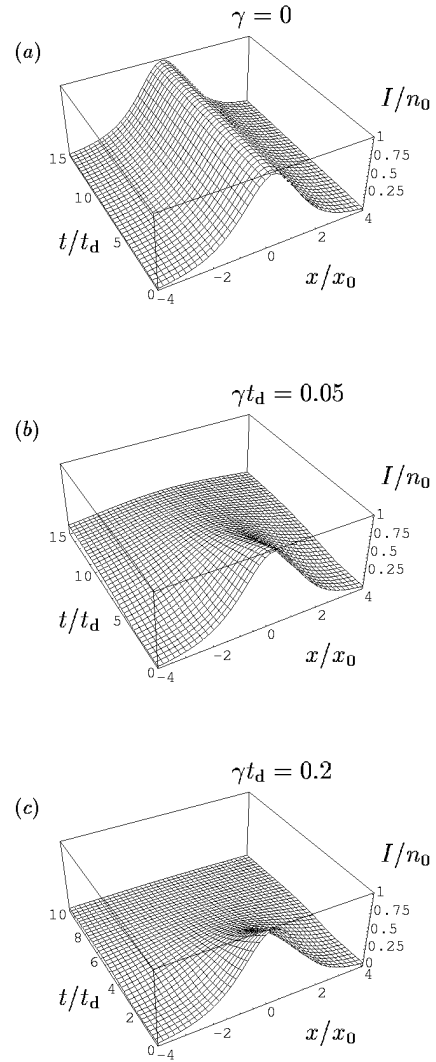,height=\heightFIG,clip=,rheight=\rheightFIG}}}
  \fi
\par
\caption{
The spatiotemporal evolution of the intensity $I$ of a
soliton is shown for various values of the damping constant $\gamma$.
}
\label{fig.int} % intensity
\end{figure}
\end{minipage}
%\fi
%\def\figINT{n}
\vspace*{2em}
\fi

\noindent
[$n_0$ $\!=$ $\!\bar{n}\Delta x/x_0$, Eq.~(\ref{eq.n0})].
The initial conditions are realized
by a multimode displaced thermal state, without entanglement
between the modes corresponding to different space points
(for displaced thermal states, see, e.g., \cite{MarianP93}).

To be more specific, we have performed the calculations
assuming a peak photon number of the initial pulse of
$\bar{n}$ $\!=$ $\!10^{9}$ and a reservoir photon number of
$N_{\rm th}$ $\!=$ $\!10^{-16}$ and setting the parameter $\delta\omega$
equal to zero. The value of $N_{\rm th}$ corresponds to a
(vacuum) carrier
wavelength of
$\lambda_{\rm c}$ $\!=$ $\!2$ $\!\!\pi$ $\!\!c$ $\!\!/\omega_{\rm c}$
$\!=$ $1.5\,\mu{\rm m}$
of the pulse in vacuum and a temperature of $T$ $\!=$ $\!300$K
[see Eq.~(\ref{Nthermisch})]. Thus the pulse is initially prepared
in a displaced thermal state that is almost a coherent state.
Assuming, e.g., losses of $\Gamma$ $\!=$ $0.3\,{\rm dB\,km}^{-1}$
[Eq.~(\ref{eq.scal.g})]
and a fiber dispersive parameter of
$D$ $\!=$ $20\,{\rm ps\,nm}^{-1}{\rm km}^{-1}$
[Eq.~(\ref{eq.k2.def})], the
(scaled) damping constant is
$\tilde\gamma$ $\!=$ $\!5.8\times 10^{-3}$
for a pulse duration of $t_0$ $\!=$ $2\,{\rm ps}$,
and $\tilde\gamma$ $\!=$ $\!1.4\times 10^{-1}$ for
$t_0$ $\!=$ $10\,{\rm ps}$,
and the dispersion lengths
are $x_{\rm d}$ $\!=$ $\!170\,{\rm m}$ and
$x_{\rm d}$ $\!=$ $\!4.2\,{\rm km}$, respectively
[see Eqs.~(\ref{eq.scal.t}) -- (\ref{eq.scal.g})].

\if\EprintServer\YES
%%\widetextAAA
%%\narrowtextAAA
%
%\def\figAUTOB{y} % auto3d
%\input{figures.inc}
%\if\figAUTOB\YES % auto3d
\begin{minipage}{\widthAAA}
\begin{figure}
\par
  \if\Figures\YES
%     \hbox{\centerline{\psfig{figure=auto3d.ps,width=\widthFIG,bbllx=75pt,bblly=80pt,bburx=340pt,bbury=360pt}}}
    \hbox{\centerline{\psfig{figure=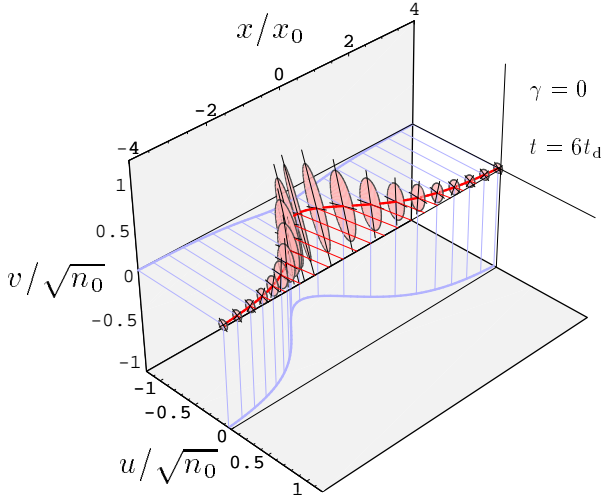,width=\widthFIG,clip=,rwidth=\rwidthFIG}}}
  \fi
\par
\caption{
The local field fluctuations of an undamped soliton
($\gamma$ $\!=$ $\!0$)
are shown for fixed propagation time $t$ and various positions $x$
in the pulse.
The uncertainty ellipses in the phase space
($s$ $\!=$ $\!0.85$)
are plotted in arbitrary units around the (solid line) complex amplitude
$\langle\hat{a}(x)\rangle$ $\!=$ $\!\langle\hat{u}(x)\rangle$
$\!+$ $\!i \langle\hat{v}(x)\rangle$.
}
\label{fig.auto3d}
\end{figure}
\end{minipage}
\vspace*{2em}
\fi
%\def\figAUTOB{n}
%\vspace*{2em}
%\def\figAUTOA{y} % auto2d
%\input{figures.inc}
%\if\figAUTOA\YES % auto2d

We have solved the equations numerically on a grid of $m$ $\!=$ $\!200$ points
with
$\Delta x$ $\!\!/x_0$ $\!=$ $\!0.1$ and absorbing boundary
conditions,
employing the Runge-Kutta method of
order $8(5,3)$ with step-size control \cite{HairerE93}.
In order to test the program,
we have performed the calculations for different values of $s$
and verified Eq.~(\ref{eq.C20.s}) to be within the numerical
round-off error.
Examples of the temporal evolution of the pulse intensity
$I$ $\!=$ $\!I(x)$ $\!\equiv$ $\!I_j$, Eq.~(\ref{eq.Ij}),
are shown in Fig.~\ref{fig.int} for
various values of $\gamma$.

%%%%%%%%%%%%%%%%%%%%%%%%%%%%%%%%%%%%%%%%%%%%%%%%%%%%%%%%%%%%%%%%%%%%%%%%%%%%%%

\subsection{Local fluctuations}
\label{Sec.local}

In the Gaussian approximation, the local field noise
(expressed in terms of $s$-ordered variances)
can be visualized in phase space by means of uncertainty ellipses
with the large $B^{1/2}$ and small $b^{1/2}$ axes
and orientation angle $\varphi$,
\begin{eqnarray}
B,b = {\textstyle\frac{1}{2}} \left[
C_{u^2}+C_{v^2}\pm\sqrt{\left(C_{u^2}-C_{v^2}\right)^2+4C_{uv}^2}
\right],
\label{eq.chi1.nn}
\end{eqnarray}
\begin{eqnarray}
2\varphi = \arg\left(C_{u^2}-C_{v^2}+2iC_{uv}\right)
\label{eq.chi1.arg}
\end{eqnarray}
(Appendix~\ref{App.nr}), where the argument $x$ is skipped for simplicity.
  From inspection of Eqs.~(\ref{eq.chi1.nn}) and (\ref{eq.chi1.arg})
together with Eq.~(\ref{eq.C20.s}) it can be seen that $b$ and $B$
depend on $s$,
\begin{eqnarray}
b \to b_s = b_0 - {\textstyle\frac{1}{4}} s 
\label{eq.bB.s}
\end{eqnarray}
(and $B$ accordingly), whereas $\varphi$ is $s$ independent.
The positions in phase space of the ellipses are determined
by the $s$-independent first-order cumulants $C_u(x)$ and $C_v(x)$.
Squeezed local fluctuations are observed if $b$ is reduced
below the vacuum level $b_{\rm vac}$,
\begin{eqnarray}
b < b_{\rm vac} \equiv {\textstyle\frac{1}{4}} (1-s)
\label{eq.sqz.cond}
\end{eqnarray}
(Appendix~\ref{App.nr}). Obviously, the property of a state
to be squeezed does not depend on $s$. For illustrational
reasons it is convenient to choose $s$ as large as possible
[otherwise the difference $b$ $\!-$ $\!(1$ $\!-$ $\!s)/4$
may not become visible].

\if\EprintServer\YES
\begin{minipage}{\widthAAA}
\begin{figure}
\par
  \if\Figures\YES
%     \hbox{\centerline{\psfig{figure=auto2d.ps,width=\widthFIG,bbllx=160pt,bblly=430pt,bburx=455pt,bbury=690pt}}}
     \hbox{\centerline{\psfig{figure=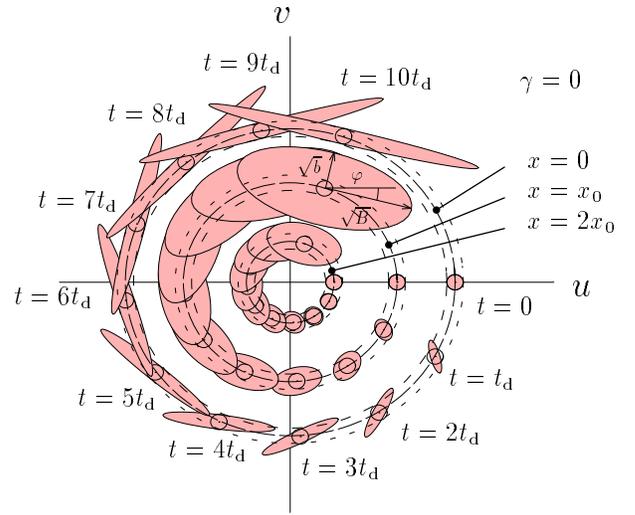,width=\widthFIG,clip=,rwidth=\rwidthFIG}}}
  \fi
\par
\caption{
The temporal evolution of the local field fluctuations
of an undamped soliton
($\gamma$ $\!=$ $\!0$) is shown for three typical positions in the pulse.
The uncertainty ellipses in the phase space
($s$ $\!=$ $\!0.85$)
are plotted in arbitrary units around the complex amplitude
$\langle\hat{a}(x)\rangle$ $\!=$ $\!\langle\hat{u}(x)\rangle$
$\!+$ $\!i \langle\hat{v}(x)\rangle$.
For comparison, the vacuum noise is indicated by circles.
}
\label{fig.auto2d}
\end{figure}
\end{minipage}
%\fi
%\def\figAUTOA{n}
\vspace*{2em}
%
%%\widetextAAA
%%\narrowtextAAA
\fi

The spatiotemporal evolution of the local field fluctuations
(together with the complex amplitude) of an undamped soliton
($\gamma$ $\!=$ $\!0$, $\chi>0$, $\omega^{(2)}<0$) 
is illustrated in Figs.~\ref{fig.auto3d}
and \ref{fig.auto2d}.
Since the uncertainty ellipses are very small
(\mbox{$\sim\!n_0^{-1/2}$})
compared with the (absolute) values of the complex amplitude
(\mbox{$\sim\!1$}), the ellipses are
plotted on an enlarged scale (of arbitrary units).
Figure \ref{fig.auto3d} represents the dependence on space of
the local fluctuations for fixed time. The vacuum-noise level can
be estimated from the uncertainties observed
sufficiently far from the pulse center ($|x|$ $\!\gtsim$ $4x_0$ in the figure),
where the field is prepared in a state close to vacuum.
The Kerr-induced self-phase modulation may be regarded as being the
main reason for the enhanced phase noise of the field in the vicinity
of the pulse center
(for continuous wave radiation, see, e.g., \cite{KitagawaM86}).
The reason for the spatial redistribution of the local fluctuations
may be seen in the dispersion that also leads to the
observed enhanced amplitude noise at the wings of the pulse
(for the influence of the dispersion on the squeezed pulses
in linear media see, e.g., \cite{SchmidtE96}).
In Fig.~\ref{fig.auto2d} the temporal evolution of the local
fluctuations is shown for various space points.
It is seen that during the pulse propagation the complex amplitude
rotates clockwise (counter-clockwise in the case 
$\chi<0$, $\omega^{(2)}>0$)
in the phase space due to the nonlinear
phase shift. The period of the rotation is close to the classical period
$4\pi t_{\rm d}$ \cite{AkhmanovSA92}, because of the large value
of $\bar{n}$. For small propagation times the field in the pulse center is
squeezed, whereas at the wings the noise exceeds
the vacuum level (indicated by the circles in the figure).
In the further course of time an increase of the noise is observed.

%\par\noindent
\if\EprintServer\YES
%\def\figBMINA{y} % bmin1d
%\input{figures.inc}
%\if\figBMINA\YES % bmin1d
\begin{minipage}{\widthAAA}
\begin{figure}
\par
  \if\Figures\YES
%     \hbox{\centerline{\psfig{figure=bmin1d.ps,width=\widthFIG,bbllx=180pt,bblly=270pt,bburx=580pt,bbury=590pt}}}
     \hbox{\centerline{\psfig{figure=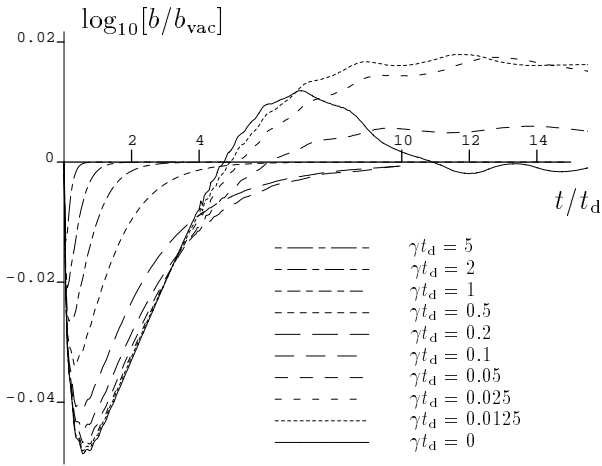,width=\widthFIG,clip=,rwidth=\rwidthFIG}}}
  \fi
\par
\caption{
The temporal evolution of the minimum-qua\-drature noise $b$ in
the center of a soliton is shown for various values of the damping constant
$\gamma$ ($s$ $\!=$ $\!0$).
}
\label{fig.bmin1d}
\end{figure}
\end{minipage}
%\fi
%\def\figBMINA{n}
\vspace*{2em}
\fi

\noindent The influence of absorption on the spatiotemporal
evolution of the minimum quadrature noise $\!\sim$ $\!b$
is shown in Figs.~\ref{fig.bmin1d} and \ref{fig.bminxt}.
  From Fig.~\ref{fig.bmin1d} that presents the temporal evolution
of $b$ in the pulse center it is seen that
absorption reduces the maximally achievable amount of squeezing.
Figure~\ref{fig.bminxt} reveals that
for small propagation times ($t$ $\!\ltsim$ $\!t_{\rm d}$)
squeezing is observed over the whole pulse, whereas
in the further course of propagation  ($t$ $\!\gtsim$ $\!t_{\rm d}$)
``noisy'' wings appear owing to dispersion-assisted noise
redistribution in the pulse.
It is also seen that absorption reduces the
effect of noise enhancement at the wings.

The local field noise can be described in terms of
squeezed thermal states, and the squeezing condition
(\ref{eq.sqz.cond}) can be rewritten as
\begin{equation}
(2n+1)e^{-2r} < 1
\label{eq.rn.sqz}
\end{equation}
(Appendix~\ref{App.nr}). Here, $n$ [Eq.~(\ref{eq.n})] is the
``thermal'' photon number associated with the local field noise,
and $r$ [Eq.~(\ref{eq.r})] is the associated squeeze parameter.
The parameter $r$ is a measure of the ``deformation''
of the uncertainty ellipses, whereas $n$ determines their ``size''.
%\noindent 
Examples of the spatiotemporal
evolution of $n$ and $r$ are shown in Figs.~\ref{fig.nxt} and
\ref{fig.rxt}, respectively.
It is seen that for small
propagation times $r$ increases more rapidly than $n$ in the pulse
center, and hence the field in the pulse center can be squeezed
after the pulse has entered the fiber.
The destructive influence of absorption is demonstrated in
Figs.~\ref{fig.nxt}$(b)$ and \ref{fig.rxt}$(b)$.

\if\EprintServer\YES
%\def\figBMINXT{y} % bminxt
%\input{figures.inc}
%\if\figBMINXT\YES % bminxt
\begin{minipage}{\widthAAA}
\begin{figure}
\par
  \if\Figures\YES
%     \hbox{\centerline{\psfig{figure=bminxt.ps,height=\heightFIG,bbllx=110pt,bblly=90pt,bburx=400pt,bbury=745pt}}}
     \hbox{\centerline{\psfig{figure=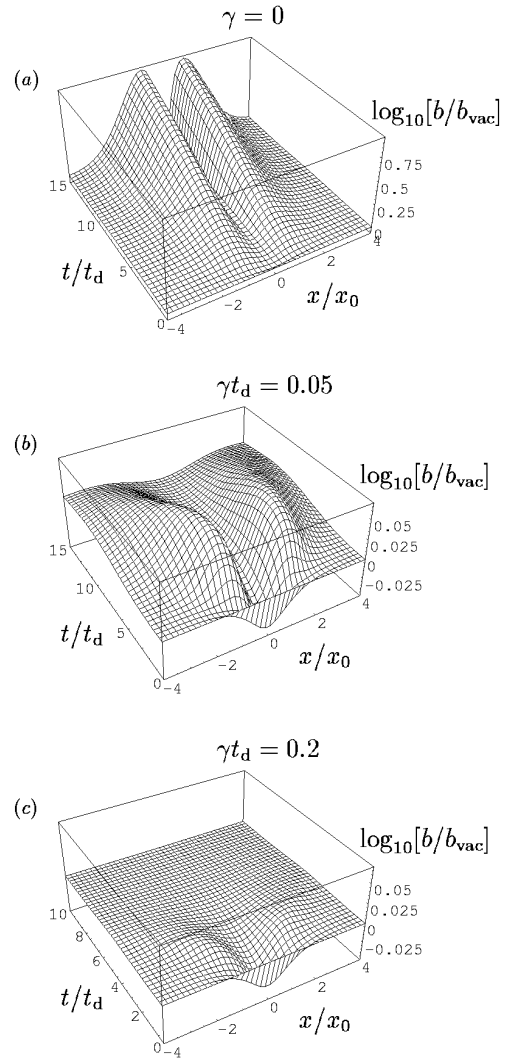,height=\heightFIG,clip=,rheight=\rheightFIG}}}
  \fi
\par
\caption{
The influence of absorption on
the spa\-tio\-tem\-poral evolution of the minimum-qua\-drature noise $b$
of a soliton is illustrated for three values of the damping constant
$\gamma$ ($s$ $\!=$ $\!0$).
}
\label{fig.bminxt}
\end{figure}
\end{minipage}
%\fi
%\def\figBMINXT{n}
\vspace*{2em}
\fi

\if\EprintServer\YES
\vfill
\newpage

\widetextAAA
\narrowtextAAA
%
%\def\figNXT{y} % nxt
%\input{figures.inc}
%\if\figNXT\YES
\begin{minipage}{\widthAAA}
\begin{figure}
\par
  \if\Figures\YES
%     \hbox{\centerline{\psfig{figure=nxt.ps,height=\heightFIG,bbllx=110pt,bblly=290pt,bburx=385pt,bbury=740pt}}}
     \hbox{\centerline{\psfig{figure=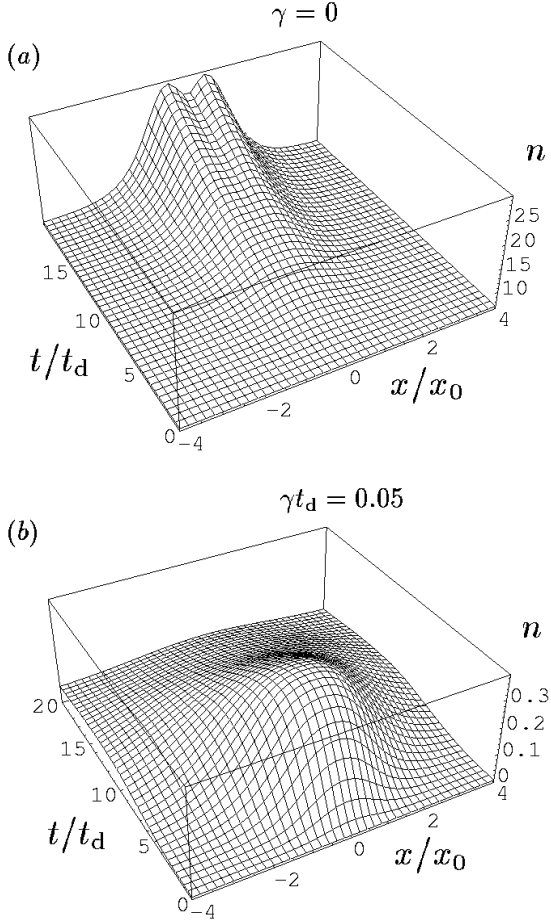,width=\widthFIGa,clip=,rwidth=\rwidthFIGa}}}
  \fi
\par
\caption{
The spatiotemporal evolution of the ``thermal'' photon number
$n$ of the squeezed thermal state associated with the
local field noise of a soliton is shown for two values
of the damping constant $\gamma$.
}
\label{fig.nxt}
\end{figure}
\end{minipage}
%\fi
%\def\figNXT{n}
%\vspace*{2em}
%

%
%\def\figRXT{y} % rxt
%\input{figures.inc}
%\if\figRXT\YES
\begin{minipage}{\widthAAA}
\begin{figure}
\par
  \if\Figures\YES
%     \hbox{\centerline{\psfig{figure=rxt.ps,height=\heightFIG,bbllx=110pt,bblly=290pt,bburx=385pt,bbury=740pt}}}
     \hbox{\centerline{\psfig{figure=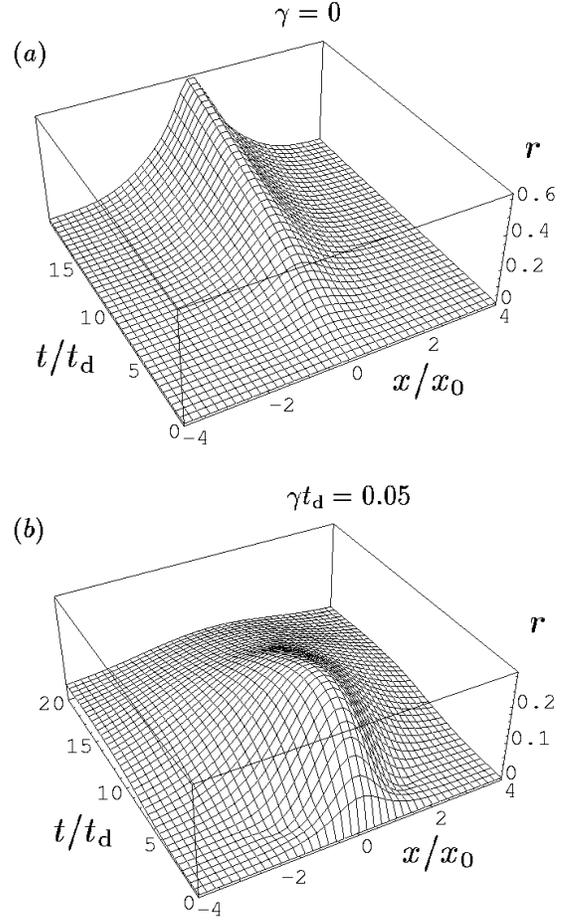,width=\widthFIGa,clip=,rwidth=\rwidthFIGa}}}
  \fi
\par
\caption{
The spatiotemporal evolution of the squeeze parameter
$r$ of the squeezed thermal state associated with the
local field noise of a soliton is shown for two values
of the damping constant $\gamma$.
}
\label{fig.rxt}
\end{figure}
\end{minipage}
%\fi
%\def\figRXT{n}
%\vspace*{2em}
%
\widetextAAA
\vspace*{-1em}
\narrowtextAAA

\fi

%%%%%%%%%%%%%%%%%%%%%%%%%%%%%%%%%%%%%%%%%%%%%%%%%%%%%%%%%%%%%%%%%%%%%%%%%%%%%%

\subsection{Squeezing spectrum}
\label{Sec.sqz.sp}

The squeezing spectrum of the (output) pulse can be measured
using balanced homodyne detection \cite{WernerMJ97}.
Neglecting outcoupling effects, the temporal profile of the
pulse after propagation through a fiber of length $L$
is described by the operator
\begin{eqnarray}
\hat{a}(L-v_{\rm gr}t,t)
= \hat{a}(-\tau v_{\rm gr},\tau+L/v_{\rm gr}),
\label{eq.a.L}
\end{eqnarray}
where $\tau$ $\!=$ $\!t$ $\!-$ $\!L/v_{\rm gr}$ is the difference
of the absolute time $t$ and the pulse arrival time $L/v_{\rm gr}$.
The approximation in Eq.~(\ref{eq.wk}) requires the temporal pulse
width $t_0$ to be small compared with the dispersion time $t_{\rm d}$,
Eq.~(\ref{eq.td}),
\begin{eqnarray}
t_0/t_{\rm d}\sim \Delta k|\omega^{(2)}|\,/\,|\omega^{(1)}|\ll 1.
\label{eq.t.weg}
\end{eqnarray}
Hence, the dependence on $\tau$ of the second argument in
Eq.~(\ref{eq.a.L}) can be ignored, and the time profile of
the pulse can be described by
\begin{eqnarray}
\hat{a}(-\tau v_{\rm gr},L/v_{\rm gr})=
\hat{a}(x,L/v_{\rm gr})\equiv\hat{a}(x),
\label{eq.a.Lneu}
\end{eqnarray}
where for notational reasons the
time coordinate $\tau$ is formally replaced by the spatial
coordinate $x$

   From Appendix~\ref{App.sqz.sp}, the squeezing spectrum $S(\omega)$
can be expressed in terms of (discrete) $s$-ordered cumulants as
\begin{equation}
S(\omega)=2
%TCIMACRO{\func{Re} }
%BeginExpansion
\mathop{\rm Re}%
%EndExpansion
\!
\left[
   F_{\rm L}(-\omega,\omega)+e^{2i\varphi}
   G_{\rm L}(-\omega,\omega)
\right]I_0^{-1},  \label{EQ.SW}
\end{equation}
where
\begin{eqnarray}
&&\lefteqn{
F_{\rm L}(\omega,\omega')=
\frac{1}{2\pi}
 \sum_{jj'} \Big[
a_{{\rm L}\,j}^{*} a_{{\rm L}\,j'}
e^{i\omega x_j+i\omega' x_{j'}}
}
\nonumber\\&&\hspace{4ex}\times\,
 \big\{
   \langle\!\langle u_j u_{j'}\rangle\!\rangle_s
   +\langle\!\langle v_j v_{j'}\rangle\!\rangle_s
   +{\textstyle\frac{1}{2}} (s-1) \delta_{jj'}
\nonumber\\&&\hspace{12ex}
   -i\left[
        \langle\!\langle u_j v_{j'}\rangle\!\rangle_s
       -\langle\!\langle v_j u_{j'}\rangle\!\rangle_s
     \right]
 \big\}\Big],
\label{eq.F.Psi}
\\
&&\vspace{0.2em}
\lefteqn{
G_{\rm L}(\omega,\omega')=
\frac{1}{2\pi}
\sum_{jj'}
\Big[
a_{{\rm L}\,j}a_{{\rm L}\,j'}
 e^{i\omega x_j+i\omega' x_{j'}}
}
\nonumber\\&&\hspace{4ex}\times\,
 \big\{
    \langle\!\langle u_j u_{j'}\rangle\!\rangle_s
   -\langle\!\langle v_j v_{j'}\rangle\!\rangle_s
\nonumber\\&&\hspace{12ex}
   -i\left[
       \langle\!\langle u_j v_{j'}\rangle\!\rangle_s
      +\langle\!\langle v_j u_{j'}\rangle\!\rangle_s
     \right]
 \big\}\Big].
\label{eq.G.Psi}
\end{eqnarray}
Here, $I_0$ is a normalization constant and
$a_{{\rm L}\,j}$
is the complex local-oscillator (discrete) amplitude,
with $\varphi$ being the relative phase between
the local oscillator and the signal.
Negative values of $S(\omega)$ signify squeezing at
frequency $\omega$.
Note that the squeezing spectrum involves
internal pulse correlations expressed in terms of the
two-argument cumulants in Eqs.~(\ref{eq.F.Psi}) and (\ref{eq.G.Psi}).
For the chosen frequency $\omega$ the minimum of $S(\omega)$,
\begin{equation}
S_{\min}(\omega) = 2 \left[
%TCIMACRO{\func{Re} }
%BeginExpansion
\mathop{\rm Re}%
%EndExpansion
F_{\rm L}(-\omega,\omega) -
| G_{\rm L}(-\omega,\omega) | \right] I_0^{-1},
\label{eq.Smin}
\end{equation}
is observed if the phase $\varphi$ is chosen such that
\begin{equation}
2\varphi =\pi-\arg\!
\left[ G_{\rm L}(-\omega,\omega) \right] .
\label{eq.opt.phase}
\end{equation}
\if\EprintServer\YES
%\def\figSSA{y} % ss1d
%\input{figures.inc}
%\if\figSSA\YES % ss1d
\begin{minipage}{\widthAAA}
\begin{figure}
\par
  \if\Figures\YES
%     \hbox{\centerline{\psfig{figure=ss1d.ps,width=\widthFIG,bbllx=110pt,bblly=300pt,bburx=540pt,bbury=570pt}}}
     \hbox{\centerline{\psfig{figure=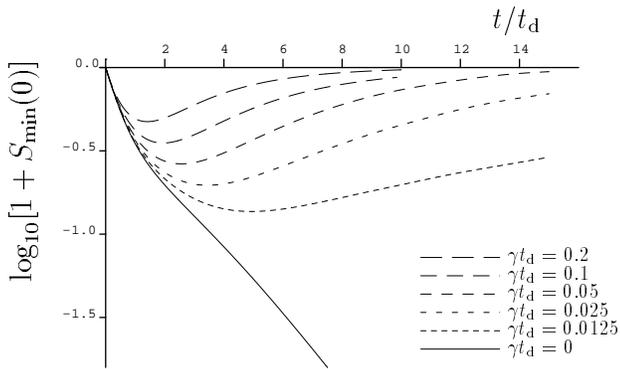,width=\widthFIG,clip=,rwidth=\rwidthFIG}}}
  \fi
\par
\caption{
The influence of absorption on the temporal evolution
of the midcomponent of the squeezing spectrum 
$S_{\min}(0)$, Eq.~(\protect\ref{EQ.SW}), is shown.
}
\label{fig.ss1d}
\end{figure}
\end{minipage}
%\fi
%\def\figSSA{n}
\vspace*{2em}
\fi

The temporal evolution of the minimum
squeezing spectrum $S_{\min}(\omega)$
is shown in Figs.~\ref{fig.ss1d} and \ref{fig.ssxt} 
($\omega_0$ $\!=$ $2/x_0$).
It is assumed that the local oscillator (LO)
pulse has the shape of the initial
fundamental-soliton pulse. Figure \ref{fig.ss1d} demonstrates the
influence of absorption on the temporal evolution of the squeezing spectrum
for $\omega$ $\!=$ $\!0$, and examples of the full squeezing spectrum
are shown in Fig.~\ref{fig.ssxt}.
At the initial stage of propagation wideband squeezing is observed
which evolves to enhanced sideband noise in the further course of time
[Figs.~\ref{fig.ssxt}($a$) and \ref{fig.ssxt}($b$)].
   From Fig.~\ref{fig.ssxt}($a$) it is seen that in the lossless case
the sidebands increase with time, whereas the midcomponent decreases
(see also Fig.~\ref{fig.ss1d}).
Absorption introduces bounds, as can be seen from
Fig.~\ref{fig.ssxt}($b$).
In particular, for sufficiently
strong absorption the sidebands do not appear at all
[Fig.~\ref{fig.ssxt}($c$)].

It should be noted that in the numerical implementation
of the $x$ integrals the value of $\Delta x$ imposes
according to the sampling theorem \cite{NumRecC94}
a condition on the upper bound of frequency:
$|\omega|$ $\!\le$ $\!\omega_{\rm max}$
$\!=$ $\!\pi(\Delta x)^{-1}$.
Further, there is a minimally resolvable frequency
difference $\Delta \omega_{\rm min}$ $\!=$
$\!2\pi(m \Delta x)^{-1}$. For $\Delta x/x_0$ $\!=$ $\!0.1$
and $m$ $\!=$ $\!200$ we find that 
$\omega_{\rm max}$ $\!=$ $\!5\pi$ $\!\omega_0$ and 
$\Delta \omega_{\rm min}$ $\!=$ $\!0.05\pi$ $\!\omega_0$.
%$\omega_{\rm max}x_0$ $\!\approx$ $\!30$ and 
%$\Delta \omega_{\rm min}x_0$ $\!\approx$ $\!0.15$.

%%%%%%%%%%%%%%%%%%%%%%%%%%%%%%%%%%%%%%%%%%%%%%%%%%%%%%%%%%%%%%%%%%%%%%%%%%%%%%

\subsection{Spectral photon-number correlations}

Recently experiments have been performed in order to measure the
correlation of the number of photons $\hat{N}(\omega)$ at different
frequencies \cite{SpaelterS98}.
A measure of the degree of correlation is the correlation coefficient
\begin{eqnarray}
\eta (\omega ,\omega')\!=\!
\frac{  \hspace{-1ex}
  \langle: \!\Delta \hat{N}(\omega) \Delta \hat{N}(\omega')\! :\rangle
}{
 \big[
  \langle \Delta \hat{N}^2(\omega) \rangle \,
  \langle \Delta \hat{N}^2(\omega')  \rangle
 \big]^{1/2}
}\,,
\label{eq.cor.coef}
\end{eqnarray}
\if\EprintServer\YES

%\def\figSSXT{y} % ssxt
%\input{figures.inc}
%\if\figSSXT\YES % ssxt
\begin{minipage}{\widthAAA}
\begin{figure}
\par
  \if\Figures\YES
%     \hbox{\centerline{\psfig{figure=ssxt.ps,height=\heightFIG,bbllx=120pt,bblly=80pt,bburx=390pt,bbury=720pt}}}
     \hbox{\centerline{\psfig{figure=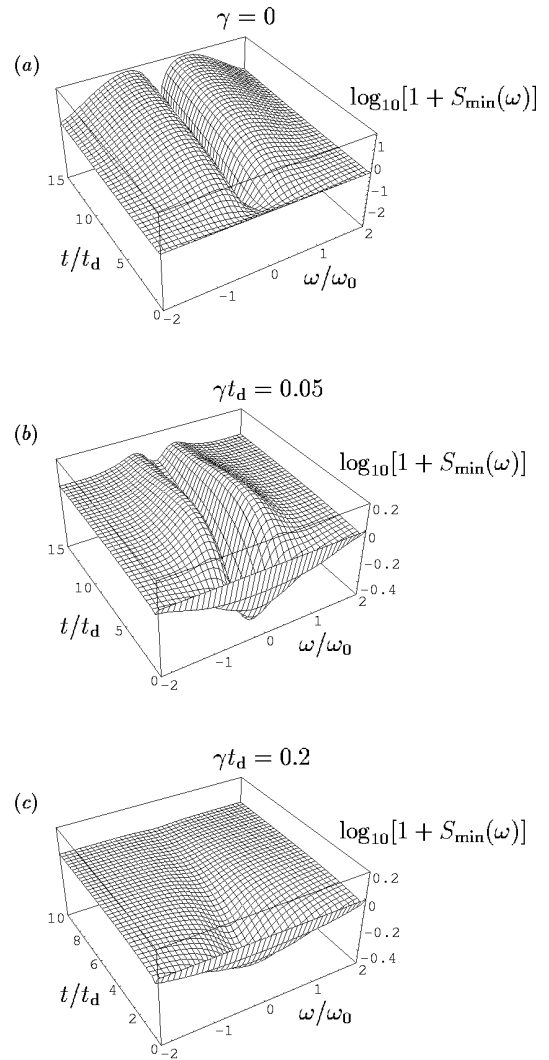,height=\heightFIG,clip=,rheight=\rheightFIG}}}
  \fi
\par
\caption{
The temporal evolution of
the squeezing spectrum $S_{\min}(\omega)$,
Eq.~(\protect\ref{EQ.SW}), is shown for various values of
the damping parameter $\gamma$.
}
\label{fig.ssxt}
\end{figure}
\end{minipage}
%\fi
%\def\figSSXT{n}

\vfill
\newpage

\widetextAAA
\narrowtextAAA
%
%\def\figNNBL{y} % nn2dl
%\input{figures.inc}
%\if\figNNBL\YES % nn2dl
\begin{minipage}{\widthAAA}
\begin{figure}
\par
  \if\Figures\YES
%     \hbox{\centerline{\psfig{figure=nn2d.ps,height=\heightFIG,bbllx=120pt,bblly=300pt,bburx=400pt,bbury=740pt}}}
     \hbox{\centerline{\psfig{figure=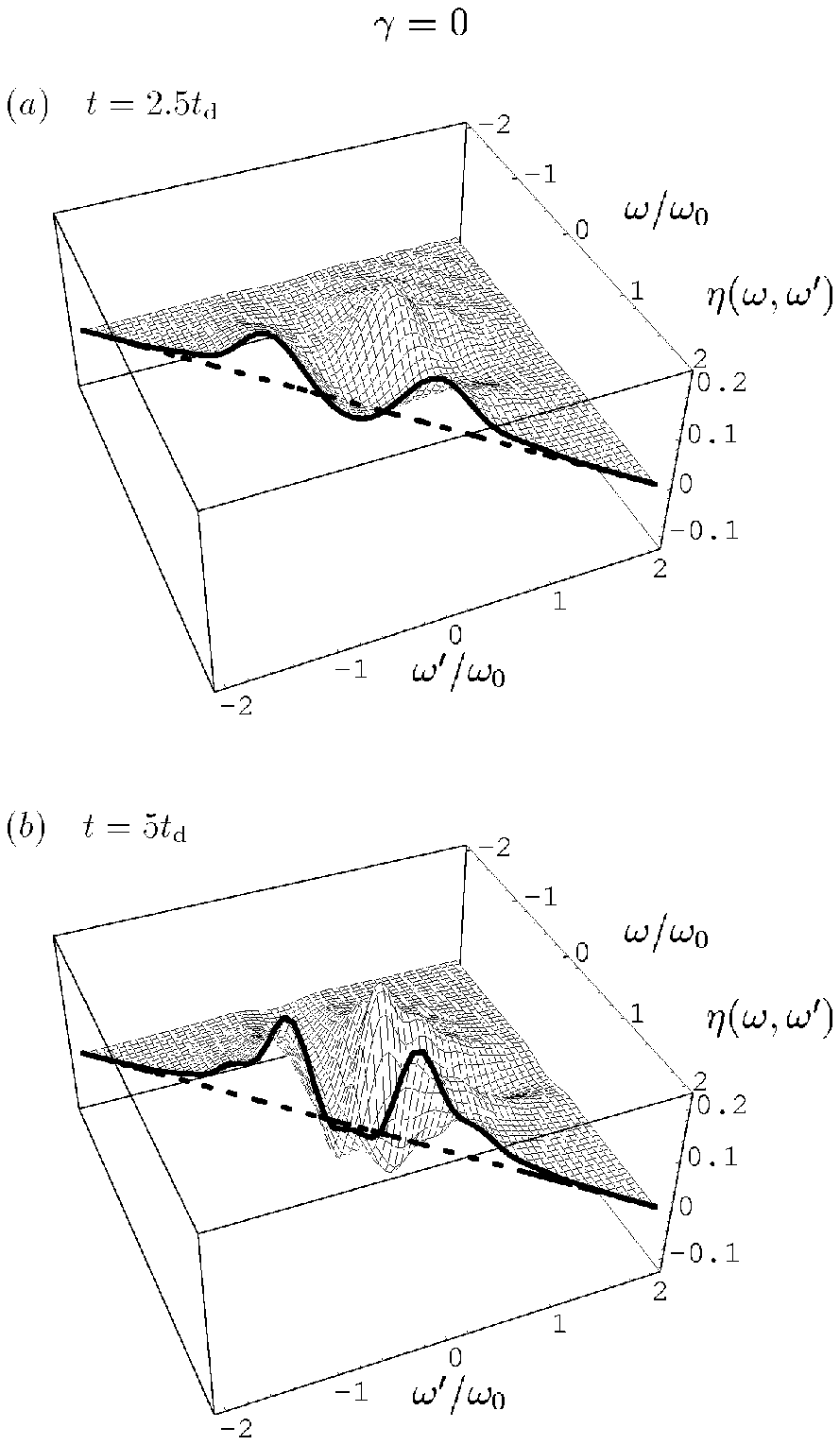,height=\heightFIG,clip=,rheight=\rheightFIG}}}
  \fi
\par
\caption{
The photon-number correlation coefficient $\eta(\omega,\omega')$,
Eq.~(\protect\ref{eq.cor.coef}), of an undamped soliton
($\gamma$ $\!=$ $\!0$) is shown for two
values of the propagation time $t$.
Note that the symmetry relation
\protect$\eta(\omega,\omega')$
\protect$\!=$
\protect$\eta(\omega',\omega)$
is valid.
}
\label{fig.nn2dl}
\end{figure}
\end{minipage}
%\fi
%\def\figNNBL{n}
%\vspace*{2em}

%\def\figNNB{y} % nn2d
%\input{figures.inc}
%\if\figNNB\YES % nn2d
\begin{minipage}{\widthAAA}
\begin{figure}
\par
  \if\Figures\YES
%     \hbox{\centerline{\psfig{figure=nn2d.ps,height=\heightFIG,bbllx=120pt,bblly=300pt,bburx=400pt,bbury=740pt}}}
     \hbox{\centerline{\psfig{figure=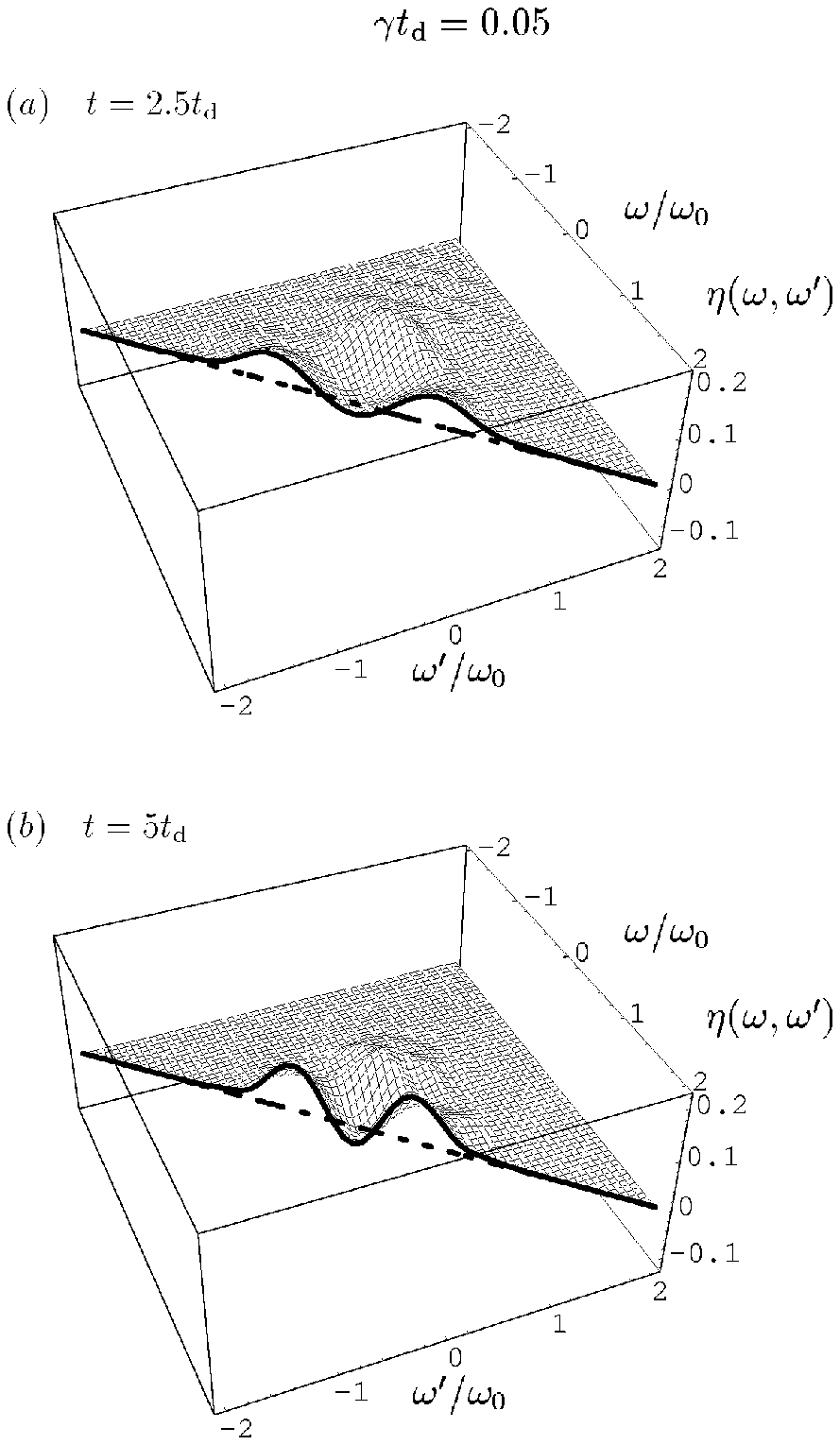,height=\heightFIG,clip=,rheight=\rheightFIG}}}
  \fi
\par
\caption{
The photon-number correlation coefficient $\eta(\omega,\omega')$,
Eq.~(\protect\ref{eq.cor.coef}), of a damped soliton
($\gamma t_{\rm d}$ $\!=$ $\!0.05$) is shown for two
values of the propagation time $t$.
Note that the symmetry relation
\protect$\eta(\omega,\omega')$
\protect$\!=$
\protect$\eta(\omega',\omega)$
is valid.
}
\label{fig.nn2d}
\end{figure}
\end{minipage}
%\fi
%\def\figNNB{n}
%\vspace*{2em}
%
\widetextAAA
\vspace*{0.5em}
\narrowtextAAA
\fi
\noindent
where $\langle : \!\Delta \hat{N}(\omega) \Delta \hat{N}(\omega')\! :\rangle$
is the normally ordered pho\-ton-number covariance, which is related to the
ordinary covariance as
\begin{eqnarray}
\lefteqn{
\langle \Delta\hat{N}(\omega) \Delta\hat{N}(\omega') \rangle
}
\nonumber\\&&\hspace{2ex}
=  \langle : \! \Delta\hat{N}(\omega) \Delta\hat{N}(\omega') \! :\rangle
+ \langle \hat{N}(\omega) \rangle
\delta(\omega,\omega')
\label{eq.cov.nn5}
\end{eqnarray}
(Appendix~\ref{Sec.nn.cov}).
The second term in Eq.~(\ref{eq.cov.nn5}), which
represents a (quantum) shot-noise contribution,
vanishes for $\omega$ $\!\neq$ $\omega'$, and hence
$\eta(\omega,\omega')$ is the ordinary correlation
coefficient. For $\omega$ $\!=$ $\omega'$ the auto\-correlation
coefficient $\eta(\omega,\omega)$ is nothing but
the (normalized) spectral photon-number variance in
normal order. Negative (positive) values of $\eta(\omega,\omega)$
correspond to sub-Poisson (super-Poisson) photon-number statistics
at chosen frequency $\omega$.

Expressing
$\langle \hat{N}(\omega)\rangle$ and
$\langle : \!\Delta \hat{N}(\omega) \Delta \hat{N}(\omega')\! :\rangle$
in terms of (discrete) $s$-ordered cumulants yields
\begin{eqnarray}
\langle \hat{N}(\omega)\rangle=
\left[ F(-\omega,\omega)+\left| E(\omega)\right| ^{2}\right] ,
\label{EQ.DN.DN1}
\end{eqnarray}
\begin{eqnarray}
\lefteqn{
\langle : \! \Delta \hat{N}(\omega)\Delta \hat{N}(\omega') \! :\rangle
= \left| F(-\omega,\omega')\right| ^{2}
 +\left| G( \omega,\omega')\right| ^{2}
}
\nonumber \\&&\hspace{1ex}
+ \, 2\func{Re}\left[ F(-\omega,\omega')E(\omega)E^{*}\!(\omega')
+ G(\omega,\omega') E^{*}\!(\omega)E^{*}\!(\omega')\right] ,
\nonumber \\ &&
\label{EQ.DN.DN}
\end{eqnarray}
where
\if\EprintServer\YES

\vfill
\newpage

\widetextAAA
\narrowtextAAA
%
%\def\figNNCL{y} % nn2dlc
%\input{figures.inc}
%\if\figNNCL\YES % nn2dlc
\begin{minipage}{\widthAAA}
\begin{figure}
\par
  \if\Figures\YES
%     \hbox{\centerline{\psfig{figure=ss2d.ps,height=\heightFIG,bbllx=120pt,bblly=300pt,bburx=400pt,bbury=740pt}}}
     \hbox{\centerline{\psfig{figure=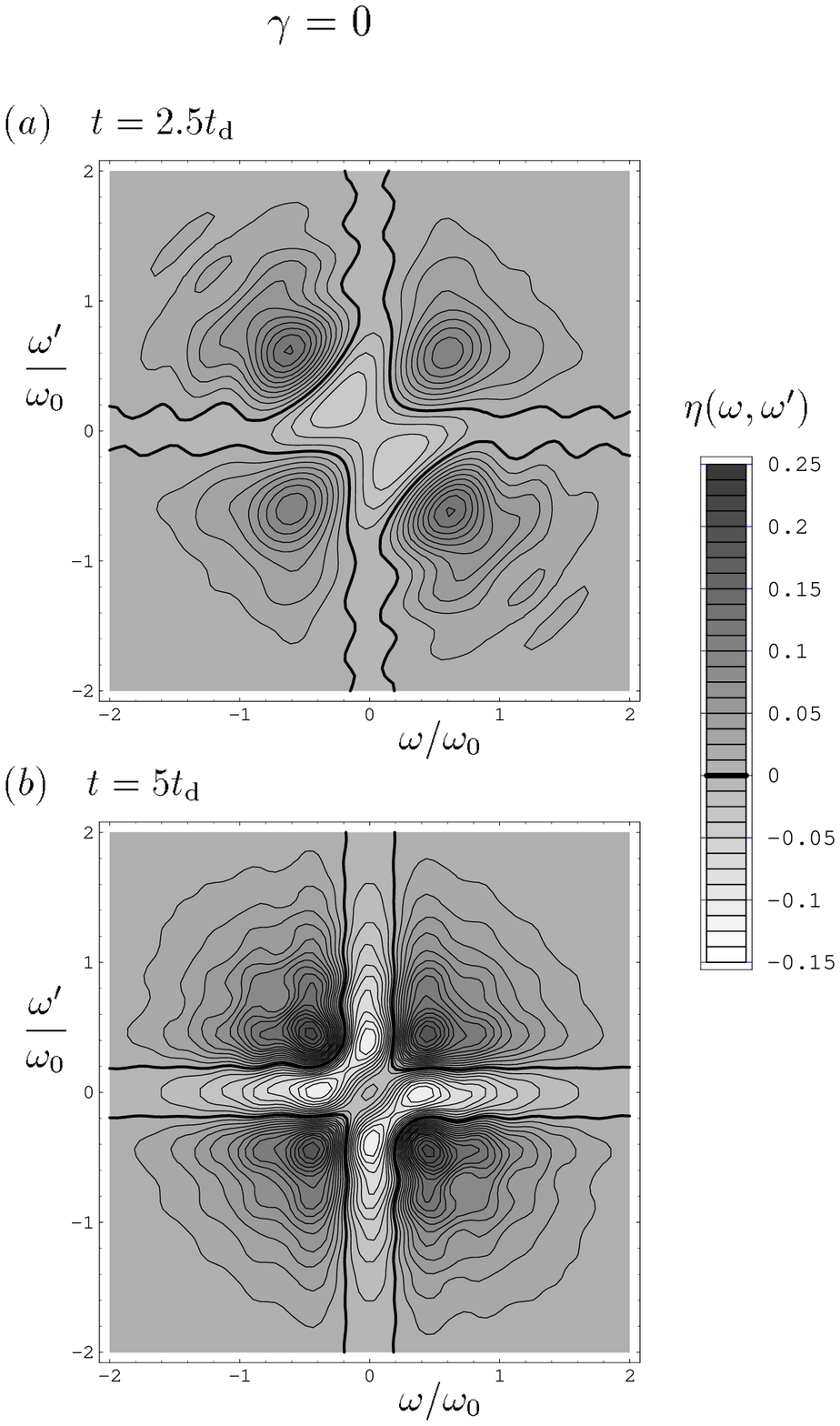,height=\heightFIG,clip=,rheight=\rheightFIG}}}
  \fi
\par
\caption{
Contour plot of
the photon-number correlation coefficient
$\eta(\omega,\omega')$, Eq.~(\protect\ref{eq.cor.coef}),
of an undamped soliton for the
same parameters as in Fig.~\protect\ref{fig.nn2dl}.
}
\label{fig.nn2dlc}
\end{figure}
\end{minipage}
%\fi
%\def\figNNCL{n}
%\vspace*{2em}

%\def\figNNC{y} % nn2dl
%\input{figures.inc}
%\if\figNNC\YES %nn2dc
\begin{minipage}{\widthAAA}
\begin{figure}
\par
  \if\Figures\YES
%     \hbox{\centerline{\psfig{figure=ss2d.ps,height=\heightFIG,bbllx=120pt,bblly=300pt,bburx=400pt,bbury=740pt}}}
     \hbox{\centerline{\psfig{figure=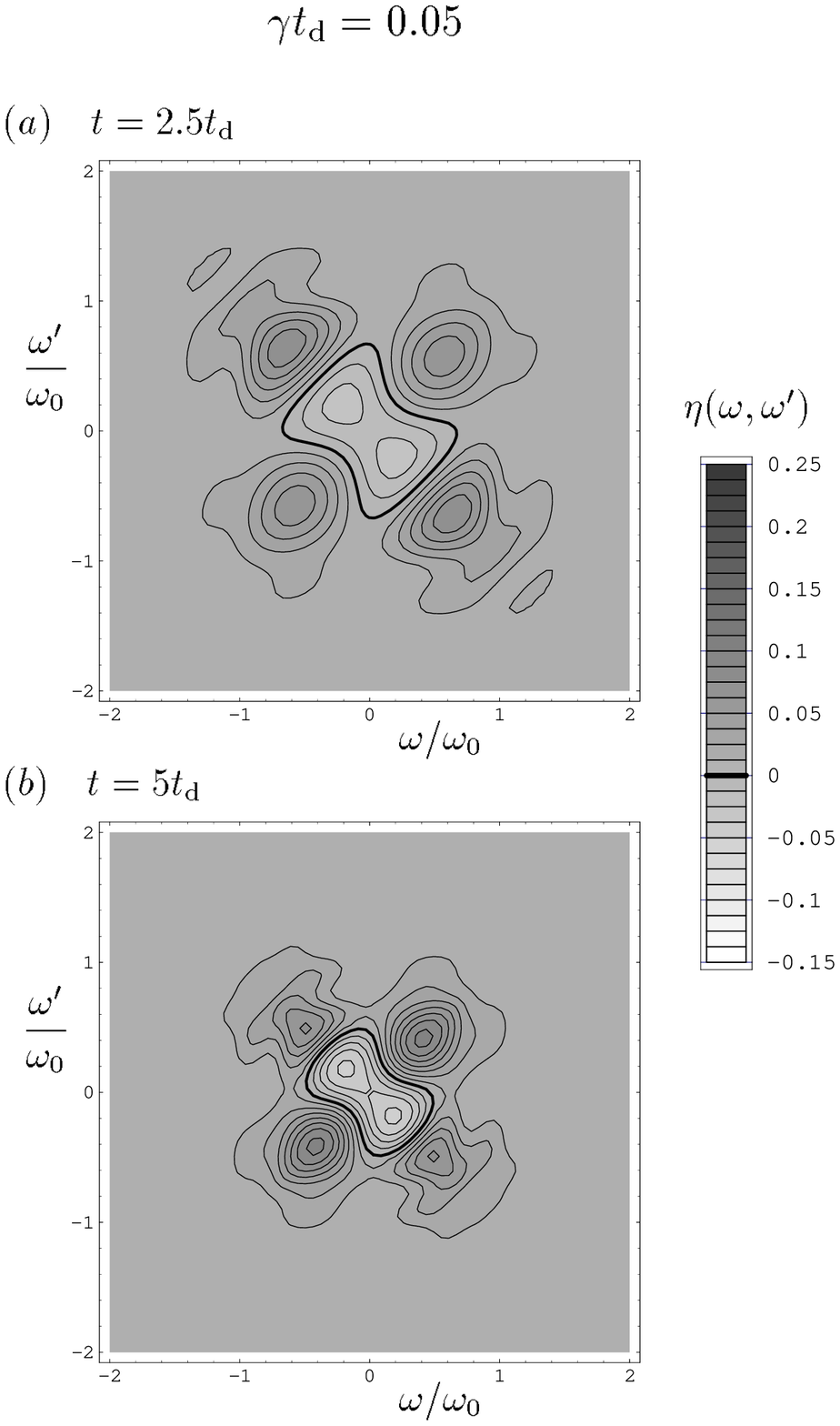,height=\heightFIG,clip=,rheight=\rheightFIG}}}
  \fi
\par
\caption{
Contour plot of
the photon-number correlation coefficient
$\eta(\omega,\omega')$, Eq.~(\protect\ref{eq.cor.coef}),
of a damped soliton for
the same parameters as in Fig.~\protect\ref{fig.nn2d}.
}
\label{fig.nn2dc}
\end{figure}
\end{minipage}
%\fi
%\def\figNNC{n}
%\vspace*{2em}
%
\widetextAAA
\vspace*{0.5em}
\narrowtextAAA
\fi
%
%
%
%where
\begin{eqnarray}
\lefteqn{
E(\omega)=
\sqrt{\frac{\Delta x \Delta\omega}{2\pi}}
 \sum_j
e^{i\omega x_j}
   \left[
      \langle\!\langle u_j \rangle\!\rangle_s
    +i\langle\!\langle v_j \rangle\!\rangle_s
   \right],
\label{eq.dn.dn2}
}
\\
\lefteqn{
F(\omega,\omega')=
\frac{\Delta x \Delta\omega}{2\pi}
 \sum_{jj'}
e^{i\omega x_j+i\omega' x_{j'}}
\big\{{\textstyle\frac{1}{2}}(s-1)\delta_{jj'}
}
\nonumber\\&&\hspace{1ex}
+\, \langle\!\langle u_j u_{j'}\rangle\!\rangle_s
   \!+\!\langle\!\langle v_j v_{j'}\rangle\!\rangle_s
   \!+\!i\left[
       \langle\!\langle u_j v_{j'}\rangle\!\rangle_s
      \!-\!\langle\!\langle v_j u_{j'}\rangle\!\rangle_s
     \right]
\big\},
\\[1ex]
\lefteqn{
G(\omega,\omega')=
\frac{\Delta x \Delta\omega}{2\pi}
 \sum_{jj'}
e^{i\omega x_j+i\omega' x_{j'}}
\big\{\langle\!\langle u_j u_{j'}\rangle\!\rangle_s
}
\nonumber\\&&\hspace{8ex}
    -\,\langle\!\langle v_j v_{j'}\rangle\!\rangle_s
   +i\left[
        \langle\!\langle u_j v_{j'}\rangle\!\rangle_s
       +\langle\!\langle v_j u_{j'}\rangle\!\rangle_s
     \right]
 \big\} ,
\label{eq.Gww}
\end{eqnarray}
with $\Delta\omega$ being the resolving frequency
(see Appendix~\ref{Sec.nn.cov}).
We have performed the calculations choosing the minimally possible
value of $\Delta \omega$ in the discretization scheme used (see the
last paragraph in Sec.~\ref{Sec.sqz.sp}).

In Figs.~\ref{fig.nn2dl} and \ref{fig.nn2dlc}
the correlation coefficient $\eta(\omega,\omega')$
of an undamped soliton is plotted for two values of the propagation time $t$.
The figures show typical features measured recently
\cite{SpaelterS98}. In particular, in the central part
of the spectrum negative correlations are observed,
whereas outside the center the frequency components
can be strongly positively correlated.
The negative values of the autocorrelation coefficient
$\eta(\omega,\omega)$ (solid lines in
Fig.~\ref{fig.nn2dl}) reveal,
in agreement with \cite{WernerMJ96}, spectral
sub-Poissonian statistics in the central part of the
spectrum followed on both sides by regions of
super-Poissonian statistics (dotted lines in
Fig.~\ref{fig.nn2dl} indicate the shot-noise level).
With increasing propagation time
the cross-correlation strength
and the super-Poissonian autocorrelation strength
are increased,
whereas the sub-Poissonian autocorrelation effect
becomes weaker
[compare Figs.~\ref{fig.nn2dl}$(a)$ and \ref{fig.nn2dlc}$(a)$ for
$t$ $\!=$ $\!2.5\,t_{\rm d}$
with Figs.~\ref{fig.nn2dl}$(b)$ and \ref{fig.nn2dlc}$(b)$ for
$t$ $\!=$ $\!5\,t_{\rm d}$ respectively].

In Figs.~\ref{fig.nn2d} and \ref{fig.nn2dc}
the correlation coefficient $\eta(\omega,\omega')$
of a damped soliton is shown for the same
values of the propagation time
as in Figs.~\ref{fig.nn2dl} and \ref{fig.nn2dlc}.
Whereas in the early stage of propagation similar
results are observed
[compare Figs.~\ref{fig.nn2dl}$(a)$ and \ref{fig.nn2dlc}$(a)$
with Figs.~\ref{fig.nn2d}$(a)$ and \ref{fig.nn2dc}$(a)$, respectively],
in the further course of time
both the negative and positive cross correlations
are smeared owing to damping
[compare Figs.~\ref{fig.nn2dl}$(b)$ and \ref{fig.nn2dlc}$(b)$
with Figs.~\ref{fig.nn2d}$(b)$ and \ref{fig.nn2dc}$(b)$, respectively].
In contrast to the cross correlations, the effect
of sub-Poissonian autocorrelations in the center of the
spectrum can be stronger for the damped soliton than for
the undamped one. This surprising effect may be
a result of damping out destructive fluctuation
interferences that occur for vanishing damping.

As mentioned above, from the theory sub-Poissonian
statistics is expected to occur in the central part of the
spectrum. Further the calculation yields a correlation
coefficient that fulfills the symmetry relation
$\eta(\omega,\omega')$ $\!=$ $\!\eta(-\omega,-\omega')$.
The two effects have not been observed in the experiment
in \cite{SpaelterS98}. A reason for the discrepancies may
be seen in phonon-scattering-induced noise which is not
considered here. It should be mentioned that for small
damping an observable sub-Poissonian effect is expected
to occur only at a rather early stage of pulse propagation
[Figs.~\ref{fig.nn2dl}$(a)$
and \ref{fig.nn2dlc}$(a)$].

%%%%%%%%%%%%%%%%%%%%%%%%%%%%%%%%%%%%%%%%%%%%%%%%%%%%%%%%%%%%%%%%%%%%%%%%%%%%%%%
%%%%%%%%%%%%%%%%%%%%%%%%%%%%%%%%%%%%%%%%%%%%%%%%%%%%%%%%%%%%%%%%%%%%%%%%%%%%%%%

\section{Summary and concluding remarks}
\label{Sec.summary}

We have studied the influence of absorption
on the quantum statistics of optical solitons in Kerr media,
introducing multivariable $s$-parametrized phase-space
distributions for describing the multimode quantum state
of a solitonlike pulse. The evolution equations for
the phase-space distributions can be tried to be solved using
cumulant-expansion techniques. When the damping is not too small,
then the quantum state can be treated in Gaussian approximation;
i.e., only cumulants up to second order are to be taken
into account. Otherwise, the evolution time must not be too large.
Starting with an undamped fundamental soliton prepared in a
nearly coherent state, we have solved the set of differential equations
in Gaussian approximation numerically. From
the solution, we have analyzed the local field fluctuations in terms
of minimum-quadrature noise and squeezed thermal states,
and we have calculated the squeezing spectrum
and the spectral photon-number correlation coefficient.

The cumulant evolution equations in the rigorous Gaussian 
approximation used in this article contain nonlinear terms 
that are disregarded in the linearization approximation. 
The obtained results show 
that for sufficiently big coherent amplitude the 
Gaussian approximation closely corresponds to the linearization 
approximation. Deviations of the exact behavior of the
quantum noise from the behavior obtained in the
linearization approximation may therefore be regarded
as being of non-Gaussian type. In this way, the cumulant method
offers a possibility of improving the theory, including in it 
higher-order cumulants --- at least third-order cumulants --- for
describing the non-Gaussian effects in a consistent way. 
 
The method can of course be used in order to study
further aspects of damped quantum soliton motion.
For example, the class of initial conditions can be extended.
In practice, the pulses which are fed into a fiber
are not of the fundamental-soliton type in general.
Further, sequences of pulses can give rise to
interactions between them.
When the losses are very small, then non-Gaussian effects
may be observable, which requires inclusion in the theory of higher
than second-order cumulants.
The model of the Kerr medium can also be improved by including
effects like Raman and Brillouin scattering \cite{CarterSJ95},
third-order dispersion \cite{SingerF92},
and frequency-dependent absorption.
In order to further specify the nonclassical features of
solitonlike pulses under realistic conditions and compare
the quantum motion with the classical one, additional properties
must be studied. With regard to the multimode structure
of solitons, the study of nonclassical internal pulse correlations
plays an important role.
Finally, application to other systems is also possible,
e.g., solitons in $\chi ^{(2)}$ media
\cite{WernerMJ97},
self-induced transparency solitons \cite{WatanabeK89},
or video pulses \cite{MaimistovAI97}.

\acknowledgments
This work was supported by the Deutsche Forschungsgemeinschaft.
We are grateful to 
S.~M.~Barnett, M.~Dakna, N.~Korolkova, and S.~Sp\"{a}lter
for valuable discussions.

%%%%%%%%%%%%%%%%%%%%%%%%%%%%%%%%%%%%%%%%%%%%%%%%%%%%%%%%%%%%%%%%%%%%%%%%%%%%%%
%%%%%%%%%%%%%%%%%%%%%%%%%%%%%%%%%%%%%%%%%%%%%%%%%%%%%%%%%%%%%%%%%%%%%%%%%%%%%%
%%%%%%%%%%%%%%%%%%%%%%%%%%%%%%%%%%%%%%%%%%%%%%%%%%%%%%%%%%%%%%%%%%%%%%%%%%%%%%

\widetextAAA

%\appendix
%\begin{appendix}
\AppendixBEG

\narrowtextAAA

\section{Scaled variables}
\label{Sec.scaled.vars}

The scaled coordinates $\tilde{x}$ and $\tilde t$ are defined by
\begin{eqnarray}
\tilde{x}  &=& \frac{x}{x_0} \, ,
\label{eq.scal.x} \\
\tilde{t} &=& \frac{t}{t_{\rm d}} \,,
\label{eq.scal.t}
\end{eqnarray}
where $x_{0}$ is the spatial pulse width, and $t_{\rm d}$ is
the dispersion time, which is related to the dispersion
length $x_{\rm d}$ as
\begin{eqnarray}
t_{\rm d} = \frac{x_{d}}{v_{\rm gr}} \,,
\label{eq.td}
\end{eqnarray}
and $x_{\rm d}$ can be given by \cite{AkhmanovSA92}
\begin{eqnarray}
x_{\rm d} = \frac{t_0^2}{|k^{(2)}|}
=\frac{x_{0}^{2}v_{\rm gr}}{|\omega^{(2)}|} \,,
\label{eq.k2.def1}
\end{eqnarray}
where
$t_{0}$ $\!=$ $\!x_{0}/v_{\rm gr}$ is the temporal pulse
width and
\begin{eqnarray}
k^{(2)} = - \frac{\omega^{(2)}}{v_{\rm gr}^3}
        =-2\pi \,\frac{c D}{\omega_{\rm c}^2}
\label{eq.k2.def}
\end{eqnarray}
[for $\omega^{(2)}$ and
$v_{\rm gr}$, see Eqs.~(\ref{eq.wk}) and (\ref{group})].
In Eq.~(\ref{eq.k2.def}), the (fiber) dispersive parameter
$D$ is introduced \cite{AkhmanovSA92}.
Further the scaled damping parameter $\tilde\gamma$ is defined by
\begin{eqnarray}
\tilde\gamma = \gamma \, t_{\rm d}
= 0.05 \cdot \ln\!10 \cdot
  \Gamma \,[{\rm dB}\,{\rm km}^{-1}]  \cdot
  x_{\rm d}\,[{\rm km}].
\label{eq.scal.g}
\end{eqnarray}
Finally the scaled amplitude operators
\begin{eqnarray}
\tilde{\hat{a}}_{j} = \frac{\hat{a}_{j}}{\sqrt{n_0}}
\label{eq.a.norm}
\end{eqnarray}
are introduced, where
\begin{eqnarray}
n_0 = \bar{n} \, \frac{\Delta x}{x_0} \,.
\label{eq.n0}
\end{eqnarray}
Here
\begin{eqnarray}
\bar{n} = \left|\frac{\omega^{(2)}{\cal A}}{\chi x_0}\right|
\label{nbar}
\end{eqnarray}
is a measure of the initial photon number of the pulse
\cite{DrummondPD87,CarterSJ95}
and is typically of the order of magnitude of $10^{8}$ -- $10^{9}$
for solitons in optical fibers \cite{DrummondPD93}.

%%%%%%%%%%%%%%%%%%%%%%%%%%%%%%%%%%%%%%%%%%%%%%%%%%%%%%%%%%%%%%%%%%%%%%%%%%%%%%

\section{Evolution equations for the second-order cumulants}
\label{Sec.2nd.cums}

Using the notation introduced in
Eq.~(\ref{eq.Cxy.def}),
the evolution equations for the second-order cumulants read as
\widetextEQ
\begin{eqnarray}
&&\partial _{t}C_{u,u}(x,x')=\left\{ \gamma \,
\left( N_{\rm th}+{\frac{1-s}{2}%
}\right) +s{\bar{\chi}}\left[ C_{u}(x)\,C_{v}(x)+C_{uv}(x)\right] \right\}
\delta (x,x')  \nonumber \\
&&\quad -2\,\gamma \,C_{u,u}(x,x')+\delta \omega
\,C_{u,v}(x,x')+\delta \omega \,C_{v,u}(x,x')
\nonumber \\
&&\quad -\frac{\omega ^{(2)}}{2\Delta x^{2}}\left[ C_{u,v}(x,x'-{%
\Delta x})-2\,C_{u,v}(x,x')+C_{u,v}(x,x'+{\Delta x}%
)\right]   \nonumber \\
&&\quad -\frac{\omega ^{(2)}}{2\Delta x^{2}}{\left[ C_{v,u}(x-{\Delta x}%
,x')-2\,C_{v,u}(x,x')+C_{v,u}({x}+{\Delta }x,x')\right] }  \nonumber \\
&&\quad +{\bar{\chi}}\Big\{ C_{u,v}(x,x')\left[ C_{u}^{2}{{%
(x')}}+3\,C_{v}^{2}{{(x')}}\right] +C_{v,u}(x,x')
\left[ C_{u}^{2}{(x)}+3\,C_{v}^{2}{(x)}\right] 
%\right.   
\nonumber \\
&&\quad \quad \quad +C_{u,v}(x,x')\left[ (s-1)+C_{u^{2}}(x')\,
+3\,C_{v^{2}}(x')\right] +C_{v,u}(x,x')\left[
(s-1)+C_{u^{2}}(x)+3\,C_{v^{2}}(x)\right]   \nonumber \\
&&\quad \quad \quad +2\,C_{u,u}(x,x')\left[
C_{u}(x)\,C_{v}(x)\,+C_{u}(x')\,C_{v}(x')
+C_{uv}(x)+C_{uv}(x')\right]   \nonumber \\
&&\quad \quad \quad +C_{v}(x)\,C_{u^{2},u}(x,x')+C_{v}(x')\,
C_{u,u^{2}}(x,x')+2\,C_{u}(x')\,C_{u,uv}(x,x')
\nonumber \\
&&\quad \quad \quad +3\,C_{v}(x')\,C_{u,v^{2}}(x,x')
+2\,C_{u}(x)\,C_{uv,u}(x,x')+3\,C_{v}(x)\,C_{v^{2},u}(x,x')  \nonumber \\
&&\quad \quad \quad 
%\left. 
+C_{u,u^{2}v}(x,x')
+C_{u,v^{3}}(x,x')+C_{u^{2}v,u}(x,x')+C_{v^{3},u}(x,x')
\Big\} ,  \label{eq.Cuu}
\end{eqnarray}
\begin{eqnarray}
&&\partial _{t}C_{u,v}(x,x')={\bar{\chi}}\left[ C_{u^{2}}^{2}{(x)}%
-C_{v^{2}}^{2}{(x)}\right] \delta (x,x')-2\,\gamma
\,C_{u,v}(x,x')+\delta \omega \,C_{v,v}(x,x')-\delta
\omega \,C_{u,u}(x,x')  \nonumber \\
&&\quad +{\frac{s\,}{2}\bar{\chi}}\left[ C_{v}^{2}{{(x)}+}C_{v^{2}}(x){-}%
C_{u}^{2}{(x)}-C_{u^{2}}(x)\right] \delta (x,x')  \nonumber \\
%&&\quad   \nonumber \\
&&\quad +{\frac{\omega ^{(2)}}{2\Delta x^{2}}\,\left[ C_{u,u}(x,x'-{%
\Delta x})-2\,C_{u,u}(x,x')+C_{u,u}(x,x'+{\Delta x}%
)\right] }  \nonumber \\
&&\quad -\frac{\omega ^{(2)}}{2\Delta x^{2}}{\,\left[ C_{v,v}(x-{\Delta x}%
,x')-2\,C_{v,v}(x,x')+C_{v,v}({x}+{\Delta }x,x')\right] }  \nonumber \\
&&\quad +{\bar{\chi}}\left\{ -C_{u,u}(x,x')\left[ 3\,C_{u}^{2}{{%
(x')}}+C_{v}^{2}{{(x')}}\right] +C_{v,v}(x,x')
\left[ C_{u}^{2}{(x)}+3\,C_{v}^{2}{(x)}\right] \right.   \nonumber \\
&&\quad \quad -C_{u,u}(x,x')\left[ \left( s-1\right)
+3\,C_{u^{2}}(x')\,+C_{v^{2}}(x')\right]
+C_{v,v}(x,x')\left[ \left( s-1\right)
+C_{u^{2}}(x)\,+3\,C_{v^{2}}(x)\right]   \nonumber \\
&&\quad \quad +2\,C_{u,v}(x,x')\,\left[
C_{u}(x)\,C_{v}(x)+C_{uv}(x)-C_{u}(x')\,C_{v}(x')
-C_{uv}(x')\right]   \nonumber \\
&&\quad \quad -C_{u}(x')\,C_{u,v^{2}}(x,x')
+2\,C_{u}(x)\,C_{uv,v}(x,x')+3\,C_{v}(x)\,C_{v^{2},v}(x,x')  \nonumber \\
&&\quad \quad \left. -C_{u,u^{3}}(x,x')-C_{u,uv^{2}}(x,x')
+C_{u^{2}v,v}(x,x')+C_{v^{3},v}(x,x')\right\} ,
\label{eq.Cuv}
\end{eqnarray}
\begin{eqnarray}
&&\partial _{t}C_{v,v}(x,x')=\left\{ \gamma \,
\left( N_{\rm th}+{\frac{1-s}{2}%
}\right) -s{\bar{\chi}}\left[ C_{u}(x)\,C_{v}(x)+C_{uv}(x)\right] \right\}
\delta (x,x')  \nonumber \\
&&\quad -2\,\gamma \,C_{v,v}(x,x')-\delta \omega
\,C_{u,v}(x,x')-\delta \omega \,C_{v,u}(x,x')
\nonumber \\
&&\quad +\frac{\omega ^{(2)}}{2\Delta x^{2}}{\,\left[ C_{u,v}(x-\Delta
x,x')-2\,C_{u,v}(x,x')+C_{u,v}({x}+{\Delta }x,x')\right] }  \nonumber \\
&&\quad +{\frac{\omega ^{(2)}}{2\Delta x^{2}}\,\left[ C_{v,u}(x,x'-\Delta x)
-2\,C_{v,u}(x,x')+C_{v,u}(x,x'+{\Delta x}%
)\right] }  \nonumber \\
&&\quad -{\bar{\chi}}\left\{ C_{u,v}(x,x')\left[ 3\,C_{u}^{2}{(x)}%
+C_{v}^{2}{(x)}\right] +C_{v,u}(x,x')\left[ 3\,C_{u}^{2}{{%
(x')}}+C_{v}^{2}{{(x')}}\right] \right.   \nonumber \\
&&\quad \quad +C_{u,v}(x,x')\left[ \left( s-1\right)
+3\,C_{u^{2}}(x)+\,C_{v^{2}}(x)\,\right] +C_{v,u}(x,x')\left[
\left( s-1\right) +3\,C_{u^{2}}(x')+\,C_{v^{2}}(x')\right]
\nonumber \\
&&\quad \quad +2\,C_{v,v}(x,x')\left[
C_{u}(x)\,C_{v}(x)\,+\,C_{u}(x')\,C_{v}(x')
+C_{uv}(x)+C_{uv}(x')\right]   \nonumber \\
&&\quad \quad +2\,C_{v}(x)\,C_{uv,v}(x,x')+2\,C_{v}(x')\,
C_{v,uv}(x,x')+C_{u}(x)\,C_{v^{2},v}(x,x')  \nonumber
\\
&&\quad \quad +3\,C_{u}(x)\,C_{u^{2},v}(x,x')+3\,C_{u}(x')\,
C_{v,u^{2}}(x,x')+C_{u}(x')\,C_{v,v^{2}}(x,x')
\nonumber \\
&&\quad \quad \left. +C_{u^{3},v}(x,x')+C_{v,u^{3}}(x,x')
+C_{v^{2}u,v}(x,x')+C_{v,v^{2}u}(x,x')\right\} .
\label{eq.Cvv}
\end{eqnarray}

\narrowtextEQ

%%%%%%%%%%%%%%%%%%%%%%%%%%%%%%%%%%%%%%%%%%%%%%%%%%%%%%%%%%%%%%%%%%%%%%%%%%%%%%

\section{Squeezed thermal noise}

\label{App.nr}
For chosen $x$ the (single-mode) $s$-parametrized characteristic
function $\chi(\underline{u},\underline{v};s)$ of the local noise
in Gaussian approximation,
\begin{eqnarray}
\chi(\underline{u},\underline{v};s)
=\exp\!\left[-{\textstyle\frac{1}{2}}\left(
     C_{u^2} \underline{u}^2
 + 2 C_{uv} \underline{u}\,\underline{v}
   + C_{v^2} \underline{v}^2
\right)\right] ,
\label{eq.chi1.C}
\end{eqnarray}
can be rewritten as
\begin{eqnarray}
\chi(\underline{u},\underline{v};s)
=\exp\!\left[- {\textstyle\frac{1}{2}} \left(
    B \underline{u}'{^2} + b \underline{v}'{^2}
\right)\right],\label{eq.chi1.Bb}
\end{eqnarray}
\begin{eqnarray}
\underline{u}'+i\underline{v}'
= (\underline{u}+i\underline{v})e^{i\varphi}\,,
\label{eq.chi1.fi}
\end{eqnarray}
where $B$, $b$, and $\varphi$, respectively, are
given in Eqs.~(\ref{eq.chi1.nn}) and (\ref{eq.chi1.arg}).
The characteristic function $\chi(\underline{u},\underline{v};s)$
corresponds to a squeezed thermal state
\begin{equation}
\hat{\rho}=\hat{S}(\xi )\,\hat{\rho}_{\rm th}(n) \,
\hat{S}^{\dagger }(\xi ),
\label{eq.nr.state}
\end{equation}
which is obtained by applying the squeeze operator
\begin{eqnarray}
\hat{S}(\xi ) &=&\exp \!\left[
{\textstyle\frac{1}{2}}\left( \xi ^{*}\hat{a}^{2}-\xi
\hat{a}^\dagger{^{\,2}}\right) \right]
\end{eqnarray}
on a thermal state $\hat{\rho}_{\rm th}(n)$ of mean photon
number $n$ (for squeezed thermal states, see, e.g.,\cite{MarianP93}).
The second-order cumulant parameters in Eq.~(\ref{eq.chi1.C})
can be shown to be related to the parameters
$\xi$ $\!=$ $\!r e^{i\theta}$ and $n$ in Eq.~(\ref{eq.nr.state}) as
\begin{eqnarray}
C_{u^2}+C_{v^2} =
\left(n + {\textstyle\frac{1}{2}} \right)\cosh(2r)
- {\textstyle\frac{1}{2}} s ,
\label{eq.th4}
\end{eqnarray}
\begin{eqnarray}
C_{u^2}-C_{v^2} + 2iC_{uv} = -
e^{i\theta} \left(n + {\textstyle\frac{1}{2}} \right)\sinh(2r).
\label{eq.th5}
\end{eqnarray}
Inverting Eqs.~(\ref{eq.th4}) and (\ref{eq.th5}), the parameters
$r$, $\theta$, and $n$ can be expressed in terms of the second-order
cumulants $C_{u^2}$, $C_{uv}$, and $C_{v^2}$ as, on recalling
Eq.~(\ref{eq.chi1.nn}),
\begin{eqnarray}
n &=&2\sqrt{\left(B+ {\textstyle\frac{1}{4}} s\right)
\left(b+ {\textstyle\frac{1}{4}} s\right)} - {\textstyle\frac{1}{2}}\,,
\label{eq.n} \\
r &=&{\textstyle\frac{1}{4}}\ln \left( \frac{B+s/4}{b+s/4}\right) ,
\label{eq.r} \\
\theta&=&\arg\left(C_{v^2}-C_{u^2}-2iC_{uv}\right) .
\end{eqnarray}
As $n$ must be non-negative, from Eq.~(\ref{eq.n}) it follows that
\begin{equation}
\sqrt{\left(B+ {\textstyle\frac{1}{4}} s\right)
\left(b+ {\textstyle\frac{1}{4}} s\right)}
\geq {\textstyle\frac{1}{4}}\,,  \label{eq.Heis.unc}
\end{equation}
which is nothing but the Heisenberg uncertainty principle [note that
according to Eqs.~(\ref{eq.C20.s}) and (\ref{eq.chi1.nn})
the  quantities $B$ $\!+$ $\!s/4$ and $b$ $\!+$ $\!s/4$
do not depend on $s$].
In the squeezing condition (\ref{eq.sqz.cond}) for the local
field noise the vacuum level $b_{\rm vac}$ can be obtained
 from Eqs.~(\ref{eq.n}), with $n$ $\!=$ $\!0$ and 
$B$ $\!=$ $\!b$ $\!\equiv$ $\!b_{\rm vac}$.
  From  Eqs.~(\ref{eq.th4}) and (\ref{eq.th5}) together with
Eq.~(\ref{eq.chi1.nn}) it is easy to show
that the squeezing condition (\ref{eq.sqz.cond}) can be
rewritten to obtain the condition (\ref{eq.rn.sqz}) in
terms of $n$ and $r$.

%%%%%%%%%%%%%%%%%%%%%%%%%%%%%%%%%%%%%%%%%%%%%%%%%%%%%%%%%%%%%%%%%%%%%%%%%%%%%%

\section{Derivation of Eq.~(\protect\ref{EQ.SW})}

\label{App.sqz.sp}
Following \cite{WernerMJ97},
we introduce the generalized detection operator
\begin{eqnarray}
\hat{X}(x) = e^{i\varphi }\hat{c}(x)+e^{-i\varphi }\hat{c}^{\dagger }(x)
\label{eq.X.def}
\end{eqnarray}
and its Fourier transform
\begin{eqnarray}
\hat{\underline X}(\omega )
&=& (2\pi)^{-1/2}\int dx\, \hat{X}(x) e^{i \omega x}\nonumber\\
&=&e^{i\varphi }\hat{\underline c}(\omega )
+e^{-i\varphi }\hat{\underline c}^{\dagger }(\omega ) ,
\end{eqnarray}
where
\begin{eqnarray}
\hat{c}(x) = \hat{a}^{\dagger }(x) \hat{a}_{\rm L}(x)
-\langle \hat{a}^{\dagger }(x)  \hat{a}_{\rm L}(x)\rangle,
\label{eq.c.def}
\end{eqnarray}
with $\varphi $ being the (adjustable) phase difference between the
LO
$\hat{a}_{\rm L}(x)$
and the signal $\hat{a}(x)$.
After some algebra we derive, on using the commutation
relation (\ref{eq.aa}),
\begin{eqnarray}
0 &\leq &\left\langle \hat{\underline X}^{\dagger }(\omega )
\hat{\underline X}(\omega
)\right\rangle =\left\langle \hat{\underline X}(-\omega )
\hat{\underline X}(\omega )\right\rangle
\nonumber \\[1ex]
&=&\int \frac{dx dx'}{2\pi}\!
\left\langle \hat{X}(x)\hat{X}(x')
\right\rangle e^{i\omega (x'-x)}  \nonumber \\[1ex]
&=&[S(\omega)+1]I_0 {\cal A}^{-2},
\end{eqnarray}
where
\begin{eqnarray}
S(\omega) = \frac{{\cal A}^2}{2\pi I_0}
\int dx dx'\,
\left\langle\! :\!\hat{X}(x)\hat{X}(x')\!:\!\right\rangle
e^{i\omega (x'-x)}
\label{eq.Sw0}
\end{eqnarray}
defines the normalized squeezing spectrum, with
\begin{eqnarray}
I_{0} = \frac{{\cal A}}{2\pi} \int dx \, \left[\big\langle
\hat{a}_{\rm L}^{\dagger}(x)\hat{a}_{\rm L}(x)
\big\rangle
\! +\!\left\langle \hat{a}^{\dagger }(x)\hat{a}%
(x)\right\rangle \right].
\end{eqnarray}
In Eq.~(\ref{eq.Sw0}) the symbol $:~:$ introduces normal order.
We assume that the LO is prepared in a coherent state
[$\hat{a}_{\rm L}(x)$ $\!\to$ $\!a_{\rm L}(x)$]
and express the normally ordered signal-field correlations
in terms of (discrete) $s$-ordered cumulants,
on recalling Eqs.~(\ref{ersetzung1}) and (\ref{ersetzung2})
and using Eq.~(\ref{eq.C20.s}).
Straightforward calculation then yields the squeezing
spectrum in the form of Eq.~(\ref{EQ.SW})
together with Eqs.~(\ref{eq.F.Psi}) and (\ref{eq.G.Psi}).

%%%%%%%%%%%%%%%%%%%%%%%%%%%%%%%%%%%%%%%%%%%%%%%%%%%%%%%%%%%%%%%%%%%%%%%%%%%%%%

\section{Derivation of Eq.~(\protect\ref{EQ.DN.DN1}) and
Eq.~(\protect\ref{EQ.DN.DN})}
\label{Sec.nn.cov}

We define the operator of the number of photons in a
sufficiently small frequency interval $(\omega,\omega+\Delta\omega)$ by
\begin{eqnarray}
\hat{N}(\omega) = \hat{\underline{a}}^\dagger(\omega)
\,\hat{\underline{a}}(\omega)\,\Delta\omega {\cal A},
\label{eq.ni.def}
\end{eqnarray}
where
\begin{eqnarray}
\hat{\underline{a}}(\omega)
= (2\pi)^{-1/2} \int dx\, \hat{a}(x) e^{i\omega x}.
\label{eq.aw.def}
\end{eqnarray}
To determine the covariance
\begin{eqnarray}
\langle \Delta \hat{N}\!(\omega)\Delta \hat{N}\!(\omega') \rangle
\!=\! \langle \hat{N}\!(\omega) \hat{N}\!(\omega') \rangle
- \langle \hat{N}\!(\omega) \rangle
  \langle \hat{N}\!(\omega') \rangle,
\label{eq.cov.nn1}
\end{eqnarray}
$\Delta \hat{N}(\omega)$ $\!=$ $\!\hat{N}(\omega)$ $\!-$
$\!\langle\hat{N}(\omega)\rangle$, the mean photon number
\begin{eqnarray}
\label{eq.cov.nn2}
\langle\hat{N}(\omega)\rangle
=  \frac{\Delta\omega{\cal A}}{2\pi} \int dx dx' \,
\left\langle\hat{a}^\dagger(x) \hat{a}(x')\right\rangle
e^{i\omega(x'-x)}
\end{eqnarray}
and the photon-number correlation
\begin{eqnarray}
\label{eq.cov.nn3}
\lefteqn{
\langle\hat{N}(\omega) \hat{N}(\omega')\rangle
= \left(\frac{\Delta\omega{\cal A}}{2\pi}\right)^2
\int \Big[ dx dx' dy dy' \,
}
\nonumber\\[1ex]&&\hspace{2ex} \times \,
\left\langle\hat{a}^\dagger(x) \hat{a}(x')
\hat{a}^\dagger(y) \hat{a}(y')\right\rangle
e^{i\omega(x'-x)}e^{i\omega'(y'-y)}
\Big]
\end{eqnarray}
must be calculated. Using the commutation relation (\ref{eq.aa}),
Eq.~(\ref{eq.cov.nn3}) can be rewritten as
\begin{eqnarray}
\langle \hat{N}(\omega) \hat{N}(\omega') \rangle
=  \langle : \! \hat{N}(\omega) \hat{N}(\omega') \! :\rangle
+ \langle \hat{N}(\omega) \rangle
\delta(\omega,\omega'),
\label{eq.cov.nn4}
\end{eqnarray}
where
\begin{eqnarray}
\delta(\omega,\omega') =
\left\{
\begin{array}{ll}
1 &  {\rm if}\ |\omega-\omega'| \leq \frac{1}{2} \Delta\omega ,
\\[1ex]
0 & {\rm otherwise.}
\end{array}
\right.
\label{eq.cov.nn6}
\end{eqnarray}
We now express the normally ordered field correlations
in the integrals to be performed in terms of (discrete)
$s$-ordered cumulants. Recalling Eqs.~(\ref{ersetzung1}) and
(\ref{ersetzung2}), applying the rule (\ref{eq.means}),
and using Eq.~(\ref{eq.C20.s}),
after some algebra we arrive at Eqs.~(\ref{EQ.DN.DN1})
and (\ref{EQ.DN.DN}) together with Eqs.~(\ref{eq.dn.dn2})
-- (\ref{eq.Gww}).

%%%%%%%%%%%%%%%%%%%%%%%%%%%%%%%%%%%%%%%%%%%%%%%%%%%%%%%%%%%

\widetextAAA

%\end{appendix}
\AppendixEND

\vspace*{2cm}

\narrowtextAAA

\vspace*{-2.5cm}

%\bibliographystyle{prsty}
%\bibliography{quant01}
%%%%%%%% insert bbl %%%%%%%%%%
%\iffalse

%\fi
%%%%%%%% end bbl %%%%%%%%%%%%%

  \if\EprintServer\YES
  \else

%\def\figINT   {y}
%\def\figAUTOB {y} % auto3D
%\def\figAUTOA {y} % auto2d
%\def\figBMINA {y} % bmin1d
%\def\figBMINXT{y} % bminxt
%\def\figNXT   {y}
%\def\figRXT   {y}
%\def\figSSA   {y} % ss1d
%\def\figSSXT  {y} % ssxt
%\def\figNNBL  {y} % nn2dl
%\def\figNNB   {y} % nn2d
%\def\figNNCL  {y} % nn2dlc
%\def\figNNC   {y} % nn2dc

%\input{figures.inc}
%%%%%%%% begin figures.inc %%%%%%%%%%%%%%%%%
%%%%%%%% end figures.inc %%%%%%%%%%%%%%%%%
  \fi

%\end{multicols}
\widetextAAA

\end{document}